\pdfoutput=1
\newcommand*{\ATLASLATEXPATH}{./}
\documentclass[cernpreprint, texlive=2016, USenglish]{\ATLASLATEXPATH atlasdoc}

\usepackage[subcaption,backend=bibtex]{\ATLASLATEXPATH atlaspackage}
\usepackage{\ATLASLATEXPATH atlasbiblatex}

\usepackage{\ATLASLATEXPATH atlascontribute}

\usepackage{\ATLASLATEXPATH atlasphysics}

\graphicspath{{./}}

\usepackage{siunitx}

\usepackage{SUSY-2016-25-PAPER-defs}

\newcommand{\AtlasCoordFootnote}{%
ATLAS uses a right-handed coordinate system with its origin at the nominal interaction point (IP)
in the center of the detector and the $z$-axis along the beam pipe.
The $x$-axis points from the IP to the center of the LHC ring,
and the $y$-axis points upwards.
Cylindrical coordinates $(r,\phi)$ are used in the transverse plane, 
$\phi$ being the azimuthal angle around the $z$-axis.
The pseudorapidity is defined in terms of the polar angle $\theta$ as $\eta = -\ln \tan(\theta/2)$.
Angular distance is measured in units of $\Delta R \equiv \sqrt{(\Delta\eta)^{2} + (\Delta\phi)^{2}}$.
Rapidity is defined by $y = \frac{1}{2}\ln[(E+p_z)/(E-p_z)]$,
where $E$ is the energy and $p_z$ is the longitudinal component of the momentum
along the beam direction. }

\AtlasTitle{\vspace{-0.5cm} Search for 
electroweak production of supersymmetric states 
in scenarios with compressed mass spectra at $\sqrt{s}=13$~\TeV{} with the ATLAS detector}

\author{The ATLAS Collaboration}

\AtlasRefCode{SUSY-2016-25}

\PreprintIdNumber{CERN-EP-2017-297}

\AtlasJournalRef{\PRD\ \textbf{97} (2018) 052010}
\AtlasDOI{10.1103/PhysRevD.97.052010}

\AtlasAbstract{%
A search for electroweak production of supersymmetric particles in scenarios with compressed
mass spectra in final states with two low-momentum leptons and missing transverse
momentum is presented. This search uses \mbox{proton--proton} collision data 
recorded by the ATLAS detector at the Large Hadron Collider in \mbox{2015--2016}, 
corresponding to 36.1~fb$^{-1}$ of integrated luminosity at $\sqrt{s}=13$~TeV. 
Events with same-flavor pairs of electrons or muons with opposite electric charge are selected. 
The data are found to be consistent with the Standard Model prediction. 
Results are interpreted using simplified models of \mbox{$R$-parity}-conserving
supersymmetry in which there is a small mass difference
between the masses of the produced supersymmetric particles and the lightest neutralino.
Exclusion limits at 95\% confidence level are set on \mbox{next-to-lightest} neutralino
masses of up to 145~GeV for Higgsino production and 175~GeV for wino
production, and slepton masses of up to 190~GeV for pair production of sleptons. 
In the compressed mass regime, the
exclusion limits extend down to mass splittings of 2.5~GeV for Higgsino
production, 2~GeV for wino production, and 1~GeV for slepton production. The results are also interpreted in
the context of a radiatively-driven natural supersymmetry model with
non-universal Higgs boson masses.}

\hypersetup{pdftitle={ATLAS document},pdfauthor={The ATLAS Collaboration}}

\begin{document}

\maketitle

%\tableofcontents

%-------------------------------------------------------------------------------
\section{Introduction}
\label{sec:intro}
Supersymmetry (SUSY)~\cite{Golfand:1971iw,Volkov:1973ix,Wess:1974tw,Wess:1974jb,Ferrara:1974pu,Salam:1974ig}
predicts new states that differ by half a unit of spin from their partner Standard Model (SM) particles,
and it offers elegant solutions to several problems in particle physics.
In the minimal supersymmetric extension to the Standard Model
(MSSM)~\cite{Fayet:1976et,Fayet:1977yc},
the SM is extended to contain two Higgs doublets,
with supersymmetric partners of the Higgs bosons called Higgsinos.  These Higgsinos mix
with the partners of the electroweak gauge bosons, the so-called winos and 
the bino, to form neutralino $\widetilde{\chi}^0_{1,2,3,4}$ and chargino
$\widetilde{\chi}^\pm_{1,2}$ mass eigenstates (subscripts indicate increasing
mass). These states are collectively referred to as electroweakinos.
In this work, the lightest neutralino $\chioz$ is assumed to be the lightest
SUSY particle (LSP) and to be stable due to $R$-parity conservation~\cite{Farrar:1978xj}, which renders it a
viable dark matter candidate~\mbox{\cite{Goldberg:1983nd,Ellis:1983ew}}.
 
Scenarios involving small mass differences between heavier SUSY particles and
the LSP are referred to as compressed scenarios, or as having compressed mass
spectra. This work considers three compressed scenarios, in which the heavier SUSY particles are
produced via electroweak interactions.
The first scenario is motivated by naturalness
arguments~\cite{Barbieri:1987fn,deCarlos:1993yy}, which suggest that the
absolute value of the Higgsino mass parameter \muh{} is near the weak
scale~\cite{Barbieri:2009ev,Papucci:2011wy}, while the magnitude of the bino and
wino mass parameters, $M_1$ and $M_2$, can be significantly
larger (such as 1~\TeV), i.e.
$\lvert\muh \rvert \ll \lvert M_1 \rvert\,, \lvert M_2 \rvert$.
This results in the three lightest electroweakino states, \chioz, \chipm, and
\chitz{} being dominated by the Higgsino component.
In this case the three lightest electroweakino masses are separated by hundreds
of~\MeV~to tens of \GeV{} depending on the composition of these mass
eigenstates, which is determined by the values of $M_1$ and
$M_2$~\cite{Han:2014kaa}.
The second scenario, motivated by dark matter coannihilation
arguments~\cite{Griest:1990kh,Edsjo:1997bg}, considers the absolute values
of the $M_1$ and $M_2$ parameters to be near the weak scale and similar in
magnitude, while the magnitude of $\muh$ is significantly larger, such that
$\lvert M_1 \rvert < \lvert M_2 \rvert \ll \lvert \muh \rvert$.
The \chipm{} and \chitz{} states are consequently wino-dominated, rendering them
nearly mass degenerate~\cite{Ibe:2012sx}, and have masses of order one to tens
of \GeV{} larger than a bino-dominated LSP.
The third scenario is also favored by such dark matter arguments, but involves the
pair production of the scalar partners of SM charged leptons (sleptons $\slepton$).
In this scenario, the sleptons have masses near the weak scale and just above the mass 
of a pure bino LSP.

Experimental constraints in these compressed scenarios are limited partly by
small electroweak production cross-sections, but also by the small momenta of
the visible decay products.
The strongest limits from previous searches are from combinations of results
from the Large Electron Positron (LEP) experiments~\cite{LEPlimits,LEPlimitsSlepton,Heister:2001nk,Heister:2002mn,Heister:2002jca,Heister:2003zk,Abdallah:2003xe,Abdallah:2003xe,Acciarri:2000wy,Achard:2003ge,Abbiendi:2003ji,Abbiendi:2002vz}.
The lower bounds on direct chargino production from these results correspond to
$m(\chiopm) > 103.5$~\GeV{} for $\Delta m(\chiopm, \chioz) > 3$~\GeV{} and
$m(\chiopm) > 92.4$~\GeV{} for smaller mass differences. For sleptons,
conservative lower limits on the mass of the scalar partner of the right-handed muon, denoted $\smuon_R$, are approximately $m(\smuon_R) \gtrsim 94.6$~\GeV{}
for mass splittings down to $\Delta m(\smuon_R, \chioz) \gtrsim 2$~\GeV. For the
scalar partner of the right-handed electron, denoted $\selectron_R$, a universal
lower bound of $m(\selectron_R) \gtrsim 73$~\GeV{} independently of $\Delta
m(\selectron_R, \chioz)$ exists. Recent phenomenological studies have proposed to probe compressed mass
spectra in the electroweak SUSY sector by using leptons with small transverse momentum,
\pt, referred to as soft
leptons~\cite{Han:2014kaa,Baer:2014kya,Giudice:2010wb,Schwaller:2013baa,Gori:2013ala,Dutta:2014jda,Han:2014aea,Barr:2015eva}.

\begin{figure}[tbp]
    \centering
    \begin{subfigure}[b]{0.40\textwidth}
      \centering
      \includegraphics[width=\columnwidth]{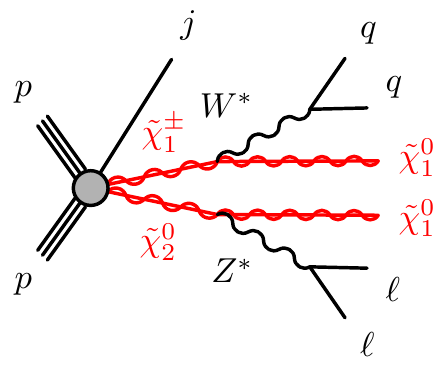}
      \caption{}
    \end{subfigure}
    \begin{subfigure}[b]{0.40\textwidth}
      \includegraphics[width=\textwidth]{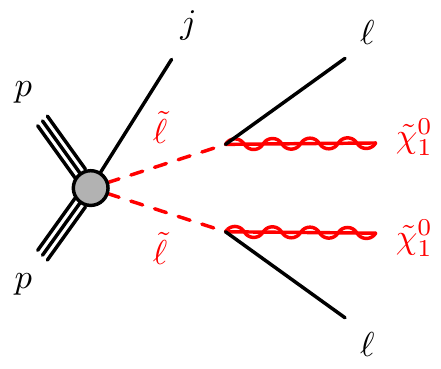}
      \caption{}
    \end{subfigure}
    \caption{\label{fig:signal_diagrams} Diagrams representing the two-lepton final state of (a) electroweakino $\chitz\chipm$ 
    and (b) slepton pair $\widetilde{\ell}\widetilde{\ell}$ production in association with a jet radiated from the initial state (labeled $j$).
    The Higgsino simplified model also considers \chitz\chioz\ and \chip\chim\ production.}
\end{figure}

\begin{figure}
    \centering
    \includegraphics[width=0.7\textwidth]{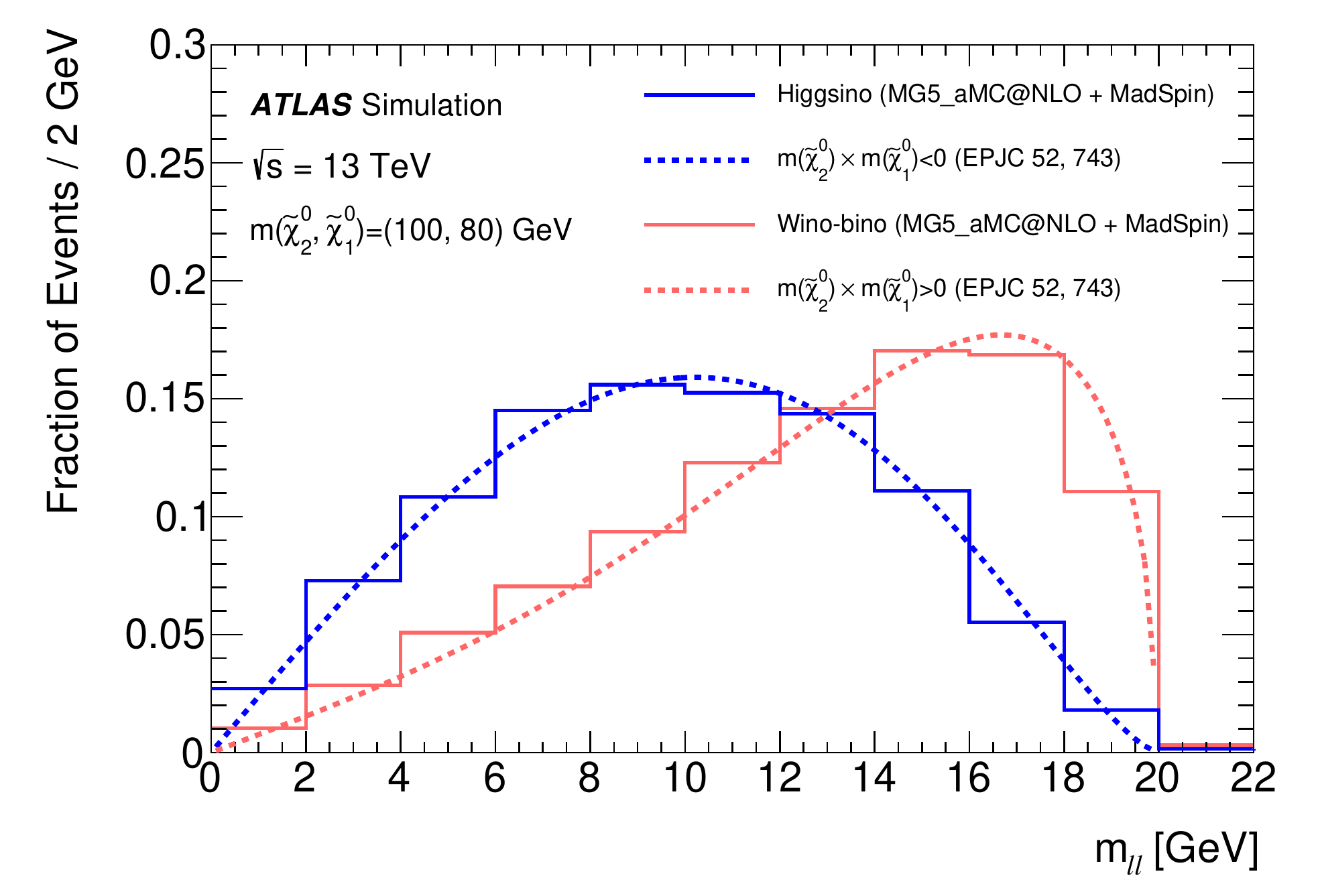}
    \caption{Dilepton invariant mass (\mll) for Higgsino and wino--bino simplified models.
      The endpoint of the \mll{} distribution is determined by the difference between the masses of the \chitz{} and \chioz.
      The results from simulation (solid) are compared with an analytic calculation of the expected lineshape (dashed)
      presented in Ref.~\cite{DeSanctis:2007yoa}, where the product of the signed
      mass eigenvalues ($m(\chitz)\times m(\chioz)$) is negative for Higgsino and positive for wino--bino scenarios.}
    \label{fig:samples:invMas}
\end{figure}

A search for electroweak production of supersymmetric
particles in compressed mass spectra scenarios with final states containing two soft same-flavor opposite-charge
leptons (electrons or muons) and a large magnitude (\met{}) of missing transverse
momentum,~\ptmiss, is presented in this paper.
The analysis uses \mbox{proton--proton ($pp$)} collision data 
collected by the ATLAS experiment from 2015 and 2016 at the Large Hadron
Collider (LHC)~\cite{Evans:2008zzb}, corresponding to 36.1~fb$^{-1}$ of integrated luminosity at $\sqrt{s}=13$~\TeV.
Figure~\ref{fig:signal_diagrams} shows schematic diagrams representing the
electroweakino and slepton pair production, as well as decays targeted in this work.
Same-flavor opposite-charge lepton pairs arise either from $\chitz$
decays via an off-shell $Z$ boson (denoted $Z^*$) or the slepton decays. The \met{} in the signal
originates from the two LSPs recoiling against hadronic initial-state radiation
(ISR).
Electroweakino signal regions are constructed using the dilepton invariant mass $m_{\ell\ell}$ as a final discriminant,
in which the signals have a kinematic endpoint given by the mass splitting of the \chitz{} and \chioz, as illustrated in
Figure~\ref{fig:samples:invMas}.  Slepton signal regions exploit a similar feature in the stransverse
mass $m_\text{T2}$~\cite{Lester:1999tx,Barr:2003rg}.
This work complements the sensitivity of existing ATLAS searches at 
$\sqrt{s} = 8$~\TeV~\cite{SUSY-2013-11,SUSY-2013-12,SUSY-2014-05,SUSY-2015-12},
which set limits on the production of winos that decay via $W$ or $Z$ bosons for mass splittings of
$\Delta m(\chipm, \chioz) \gtrsim 35$~\GeV,
and $\Delta m(\slepton, \chioz) \gtrsim 55~\GeV$ for slepton production.
Similar searches have been reported by the CMS
Collaboration at $\sqrt{s}=8$~$\TeV$~\cite{CMS-SUS-13-006,CMS-SUS-14-021} 
and at $\sqrt{s}=13~\TeV$~\cite{CMS-SUS-16-039}, which probe winos decaying via $W$ or $Z$ bosons for mass splittings
$\Delta m(\chipm, \chioz) \gtrsim 23$~\GeV.

This paper has the following structure. After a brief description of the ATLAS
detector in Section~\ref{sec:detector}, the data and Monte Carlo samples used
are detailed in Section~\ref{sec:samples}.
Sections~\ref{sec:selection} and~\ref{sec:SRs} present the event
reconstruction and the signal region selections. The background estimation
and the systematic uncertainties are discussed in Sections~\ref{sec:background}
and~\ref{sec:systematics}, respectively. Finally, the results and their
interpretation are reported in Section~\ref{sec:results} before Section~\ref{sec:conclusion} summarizes the
conclusions.

%-------------------------------------------------------------------------------
\section{ATLAS detector}
\label{sec:detector}
The ATLAS experiment~\cite{PERF-2007-01} is a general-purpose particle detector
that surrounds the interaction point with nearly $4\pi$ solid angle
coverage.\footnote{\AtlasCoordFootnote}
It comprises an inner detector, calorimeter systems, and a muon
spectrometer.
The inner detector provides precision tracking of charged particles in the pseudorapidity
region $|\eta| < 2.5$, consisting of pixel and microstrip silicon subsystems within a
transition radiation tracker. 
The innermost pixel detector layer, the insertable B-layer~\cite{Capeans:1291633}, was
added for $\sqrt{s} = 13$~TeV data-taking to improve tracking performance. 
The inner detector is immersed in a 2~T axial magnetic field provided by a superconducting
solenoid.
High-granularity lead/liquid-argon electromagnetic sampling calorimeters are
used for $|\eta| < 3.2$. Hadronic energy deposits are measured in a
steel/scintillator tile barrel calorimeter in the $|\eta| < 1.7$ region. Forward
calorimeters cover the region $1.5 <|\eta|<4.9$ for both the
electromagnetic and hadronic measurements.
The muon spectrometer comprises trigger and high-precision tracking chambers spanning $|\eta| <
2.4$ and $|\eta|<2.7$, respectively, and by three large superconducting
toroidal magnets.
Events of interest are selected using a \mbox{two-level} trigger
system~\cite{TRIG-2016-01}, consisting of a first-level trigger implemented in
hardware, which is followed by a software-based high-level trigger.
%-------------------------------------------------------------------------------

%-------------------------------------------------------------------------------
\section{Collision data and simulated event samples}
\label{sec:samples}
Searches presented here use $pp$ collision data at $\sqrt{s}=13$~TeV from the
LHC, collected by the ATLAS detector in 2015 and 2016. Events were selected using
triggers requiring large \met with run-period-dependent thresholds of 70 to 110 GeV at the trigger
level. These triggers are $>$95\% efficient for events with an
offline-reconstructed \met{} greater than 200~\GeV. The data sample corresponds to an integrated luminosity of 36.1~fb$^{-1}$ with an
uncertainty of 2.1\%, derived using methods similar to those described in
Ref.~\cite{DAPR-2013-01}.  
The average number of $pp$ interactions per bunch crossing
was 13.5 in 2015 and 25 in 2016.

Samples of Monte Carlo (MC) simulated events are used to model both the signal 
and specific processes of the SM background. 
For the SUSY signals, two sets of simplified
models~\cite{Alwall:2008ve,Alwall:2008ag,Alves:2011wf} are used to guide the
design of the analysis: one based on direct production of Higgsino states
(referred to as the Higgsino model), 
and the other a model involving pair production of sleptons 
which decay to a pure bino LSP.
For the interpretation of the results of the analysis, two additional
scenarios are considered: a simplified model assuming the production of
wino-dominated electroweakinos decaying to a bino LSP (referred to as the
wino--bino model), and a full radiatively-driven SUSY model based on
non-universal Higgs boson masses with two extra parameters
(NUHM2)~\cite{Ellis:2002wv,Ellis:2002iu}.
In all the models considered, 
the produced electroweakinos or sleptons are assumed to decay promptly.

The Higgsino simplified model includes the production of \chitz\chiopm,
\chitz\chioz\ and \chiop\chiom.
The $\chioz$ and $\chitz$ masses were varied, while the $\chiopm$
masses were set to $m(\chiopm) = \frac{1}{2}[m(\chioz) + m(\chitz)]$. 
The mass splittings of pure Higgsinos are generated by radiative corrections, and are of the order of hundreds of \MeV~\cite{Thomas:1998wy},
with larger mass splittings requiring some mixing with wino or bino states.
However, in this simplified model, the calculated cross-sections assume electroweakino mixing matrices corresponding to pure-Higgsino
$\chioz, \chiopm, \chitz$ states for all mass combinations.  
The search for electroweakinos exploits a kinematic endpoint in the dilepton invariant mass distribution, where the lepton pair is
produced in the decay chain $\chitz\to Z^*\chioz, Z^*\to\ell^+\ell^-$.   Therefore, processes that include production of a \chitz{} neutralino are
most relevant for this search, while \chiop\chiom{} production contributes
little to the overall sensitivity.
Example values of cross-sections for $m(\chitz) = 110$~\GeV{}
and $m(\chioz) = 100$~\GeV{} are 
  $4.3 \pm 0.1$~pb for \chitz\chiopm\ production and 
$2.73 \pm 0.07$~pb for \chitz\chioz\ production.
The branching ratios for $\chitz\to Z^*\chioz$ and $\chiopm \to W^* \chioz$ were fixed to 100\%.
The $Z^*\to\ell^+\ell^-$ branching ratios 
depend on the mass splittings and 
were computed using \textsc{SUSY-HIT}
v1.5b~\cite{Djouadi:2006bz}, which accounts for finite $b$-quark and
$\tau$-lepton masses. At $\Delta m(\chitz,\chioz)=60~\GeV$ the branching ratios
for $Z^*\to e^+e^-$ and $Z^*\to \mu^+\mu^-$ are approximately 3.5\%, while in
the compressed scenario at $\Delta m(\chitz,\chioz)=2~\GeV$ they increase to 5.1\%
and 4.9\%, respectively, as the $Z^*$ mass falls below the threshold needed to produce pairs of heavy quarks or $\tau$ leptons.
The branching ratios for $W^*\to e\nu$ and $W^*\to\mu\nu$ also depend on the mass splitting, and increases from 11\% for large $\Delta m(\chiopm,\chioz)$ to 20\% for $\Delta m(\chiopm,\chioz)<3~\GeV$.  
Events were generated at
leading order with \MGatNLO~v2.4.2~\cite{Alwall:2014hca} using the NNPDF23LO PDF set~\cite{Ball:2012cx} 
with up to two extra partons in the matrix element (ME).
The electroweakinos were decayed using
\MADSPIN~\cite{Artoisenet:2012st}, 
and were required to produce at least two leptons ($e$, $\mu$) in the final state,
including those from decays of $\tau$-leptons.
The resulting events were interfaced with \PYTHIA~v8.186~\cite{Sjostrand:2007gs}
using the A14 set of tuned parameters (tune)~\cite{ATL-PHYS-PUB-2014-021} to model the parton shower (PS),
hadronization and underlying event.
The ME-PS matching was performed using the CKKW-L scheme~\cite{Lonnblad:2011xx}
with the merging scale set to 15~\GeV.

The wino--bino simplified model considers \chitz\chiopm{} production, 
where the mass of the \chitz{} is assumed to be equal to that of the \chiopm.
The generator configuration as well as the decay branching ratios are consistent with those
for the Higgsino samples. Pure wino production cross-sections are used for this
model. An example value of the \chitz\chiopm\ production cross-section for
$m(\chitz,\chiopm) = 110$~\GeV{} is $16.0\pm 0.5$~pb.
The composition of the mass eigenstates differs between the wino--bino and Higgsino models.
This results in different invariant mass spectra of the two leptons originating from the virtual $Z^*$
boson in the \chitz{} to \chioz{} decay. 
The different spectra are
illustrated in Figure~\ref{fig:samples:invMas}, 
where the leptonic decays modeled by \MADSPIN{} are found
to be in good agreement with theoretical predictions 
that depend on the relative sign of the \chioz{} and \chitz{} mass parameters~\cite{DeSanctis:2007yoa},
which differs between the Higgsino and wino--bino models.

The slepton simplified model considers direct pair production
of the selectron $\selectron_{L,R}$ and smuon $\smuon_{L,R}$,
where the subscripts $L,R$ denote the left- or right-handed chirality of the partner electron or muon.
The four sleptons are assumed to be mass
degenerate, i.e.\ $m(\selectron_L) = m(\selectron_R) = m(\smuon_L) =
m(\smuon_R)$. An example
value of the slepton production cross-section for $m(\slepton_{L,R}) =
110$~\GeV{} is \mbox{$0.55\pm 0.01$~pb}.
The sleptons decay with a 100\% branching ratio into the corresponding SM partner lepton and the \chioz\ neutralino.
Events were generated at tree level using
\MGatNLO~v2.2.3 and the NNPDF23LO PDF set with up to two additional partons in the matrix element, and interfaced with \PYTHIA~v8.186
using the \mbox{CKKW-L} prescription for ME-PS matching.
The merging scale was set to one quarter of the slepton mass.
 
Higgsino, wino--bino, and slepton samples are scaled to signal cross sections calculated
at \mbox{next-to-leading} order (NLO) in the strong coupling,
and at next-to-leading-logarithm (NLL) accuracy for soft-gluon resummation,
using \textsc{Resummino}~v1.0.7~\cite{Fuks:2012qx,Fuks:2013vua,Fuks:2013lya}.
The nominal cross-section and its uncertainty
are taken from an envelope of \mbox{cross-section} predictions using different 
parton distribution function (PDF) sets and factorization and
renormalization scales, as described in Ref.~\cite{Borschensky:2014cia}.

In the NUHM2 model, the masses of the Higgs doublets that couple to the up-type and down-type quarks, 
$m_{H_{u}}$ and $m_{H_{d}}$ respectively, 
are allowed to differ
from the universal scalar masses $m_{0}$ at the grand unification scale.
The parameters of the model were fixed to the following values:
$m_{0} = 5$ TeV; the pseudoscalar Higgs boson mass $m_{A} = 1$ TeV; the
trilinear SUSY breaking parameter $A_{0} = -1.6$ $m_{0}$; the ratio of the Higgs
field vacuum expectation values $\tan\beta = 15$; and the Higgsino mass
parameter $\mu = 150$~\GeV. This choice of parameters is based on Ref.~\cite{Baer:2013xua},
which leads to a radiatively-driven natural SUSY model with low fine-tuning,
featuring: decoupled heavier Higgs bosons; a light Higgs boson with a mass of 125~\GeV{} and 
couplings like those in the SM;
colored SUSY particles with masses of the order of a few \TeV; and Higgsino-like light
electroweakinos with masses around the value of $\mu$.
The mass spectra and decay branching ratios were calculated using \textsc{Isajet}~v7.84~\cite{Baer:1999sp}.
The universal gaugino mass $m_{1/2}$ is the free
parameter in the model, and has values between 350 and 800~$\GeV$ in different event
samples. 
This parameter primarily controls the $\chitz$--$\chioz$ mass splitting, 
for example 
  $m(\chitz, \chioz) = (161, 123)~\GeV$ for $m_{1/2} = 400~\GeV$
  and
  $m(\chitz, \chioz) = (159, 141)~\GeV$ for $m_{1/2} = 700~\GeV$.
The NUHM2 phenomenology relevant to this analysis is similar to that of
the Higgsino simplified model described above, and samples of simulated
\chitz\chioz{} and \chitz\chiopm{} events were therefore generated with the same
generator configuration as the Higgsino samples, but with mass spectra, cross-sections, 
and branching ratios determined by the NUHM2 model parameters. The
cross-sections were calculated to NLO in the strong coupling constant using
\textsc{Prospino}~v2.1~\cite{Beenakker:1999xh}. They are in agreement with the
NLO calculations matched to resummation at NLL accuracy within $\sim$2\%. An
example value of the \chitz\chiopm{} production cross-section at $m_{1/2} = 700~\GeV$, corresponding to a \chitz{} mass
of 159~\GeV{} and a \chiopm~mass of 155~\GeV, is $1.07\pm 0.05$~pb.

For the SM background processes, \SHERPA versions 2.1.1, 2.2.1, and
2.2.2~\cite{Gleisberg:2008ta} were used to generate \mbox{$Z^{(*)}/\gamma^*$ +
jets}, diboson, and triboson events. Depending on the process, matrix elements
were calculated for up to two partons at NLO and up to four partons at LO using
Comix~\cite{Gleisberg:2008fv} and OpenLoops~\cite{Cascioli:2011va}, and merged
with the \SHERPA parton shower~\cite{Schumann:2007mg} according to the ME+PS@NLO
prescription~\cite{Hoeche:2012yf}. The $Z^{(*)}/\gamma^*$ + jets and diboson
samples provide coverage of dilepton invariant masses down to 0.5~\GeV{} for
$Z^{(*)}/\gamma^*\to e^+e^-/\mu^+\mu^-$, and 3.8~\GeV{} for
$Z^{(*)}/\gamma^*\to\tau^+\tau^-$.
\POWHEGBOX v1 and v2~\cite{Alioli:2010xd,Alioli:2008tz,Nason:2009ai} interfaced to
\PYTHIA 6.428 with the \textsc{Perugia}2012 tune~\cite{Skands:2010ak} were used
to simulate \ttbar\ and single-top production at NLO in the matrix element.
\POWHEGBOX v2 was also used with \PYTHIA 8.186 to simulate Higgs boson production.
\MGatNLO v2.2.2 with \PYTHIA versions 6.428 or 8.186 and the ATLAS A14 tune was
used to simulate production of a Higgs boson in association with a $W$ or $Z$
boson, as well as events containing \ttbar\ and one or more electroweak bosons.
These processes were generated at NLO in the matrix element except for $\ttbar +
WW/\ttbar$,~$t + \ttbar$, and $t + Z$, which were generated at LO. 
Table~\ref{tab:MCconfig} summarizes the generator configurations of the matrix
element and parton shower programs, the PDF sets, and the cross-section
calculations used for normalization. Further details about the generator
settings used for the above described processes can also be found in
Refs.~\cite{ATL-PHYS-PUB-2016-004,ATL-PHYS-PUB-2017-005,ATL-PHYS-PUB-2017-006,ATL-PHYS-PUB-2016-005}.

\begin{table}[]
\centering\small
\caption{Simulated samples of Standard Model background processes.  The PDF set refers to that used for the matrix element.}
\vspace{0.2cm}
\begin{tabular}{l@{\hskip9pt}l@{\hskip9pt}l@{\hskip9pt}l@{\hskip9pt}l@{\hskip9pt}l}
\toprule
Process                & Matrix element  & Parton shower & PDF set                          & Cross-section \\
\midrule
$Z^{(*)}/\gamma^*$ + jets    & \multicolumn{2}{c}{\SHERPA 2.2.1}            & NNPDF 3.0 NNLO~\cite{Ball:2014uwa}  & NNLO~\cite{Catani:2009sm} \\
Diboson                & \multicolumn{2}{c}{\SHERPA 2.1.1 / 2.2.1 / 2.2.2} & NNPDF 3.0 NNLO                      & Generator NLO \\
Triboson               & \multicolumn{2}{c}{\SHERPA 2.2.1}                 & NNPDF 3.0 NNLO                      & Generator LO, NLO \\
\midrule
\ttbar                 & \POWHEGBOX v2 & \PYTHIA 6.428 & NLO CT10~\cite{Lai:2010vv}       & NNLO+NNLL~\cite{Cacciari:2011hy,Czakon:2012pz,Czakon:2012zr,Czakon:2013goa} \\
$t$ ($s$-channel)      & \POWHEGBOX v1 & \PYTHIA 6.428 & NLO CT10                         & NNLO+NNLL~\cite{Kidonakis:2010tc}\\
$t$ ($t$-channel)      & \POWHEGBOX v1 & \PYTHIA 6.428 & NLO CT10f4                       & NNLO+NNLL~\cite{Kidonakis:2011wy,Frederix:2012dh}\\
$t + W$                & \POWHEGBOX v1 & \PYTHIA 6.428 & NLO CT10                         & NNLO+NNLL~\cite{Kidonakis:2010ux}\\
\midrule
$h(\to \ell\ell, WW)$  & \POWHEGBOX v2 & \PYTHIA 8.186 & NLO CTEQ6L1~\cite{Pumplin:2002vw}& NLO~\cite{Dittmaier:2012vm}\\
$h + W / Z$            & \MGatNLO 2.2.2 & \PYTHIA 8.186 & NNPDF 2.3 LO    & NLO~\cite{Dittmaier:2012vm}\\
\midrule
$\ttbar + W/Z/\gamma^*$& \MGatNLO 2.3.3 & \PYTHIA 8.186 & NNPDF 3.0 LO    & NLO~\cite{Alwall:2014hca}\\
$\ttbar + WW/\ttbar$   & \MGatNLO 2.2.2 & \PYTHIA 8.186 & NNPDF 2.3 LO    & NLO~\cite{Alwall:2014hca}\\
$t + Z$                & \MGatNLO 2.2.1 & \PYTHIA 6.428 & NNPDF 2.3 LO    & LO~\cite{Alwall:2014hca}\\
$t + WZ$               & \MGatNLO 2.3.2 & \PYTHIA 8.186 & NNPDF 2.3 LO    & NLO~\cite{Alwall:2014hca}\\
$t + \ttbar$           & \MGatNLO 2.2.2 & \PYTHIA 8.186 & NNPDF 2.3 LO    & LO~\cite{Alwall:2014hca}\\
\bottomrule
\end{tabular}
\label{tab:MCconfig}
\end{table}

To simulate the effects of additional $pp$ collisions, 
referred to as pileup,
additional interactions were generated using the soft QCD processes of \PYTHIA~8.186 with the A2 tune~\cite{ATL-PHYS-PUB-2012-003} and the
MSTW2008LO PDF set~\cite{Martin:2009iq}, and were overlaid onto each simulated hard-scatter event. The MC
samples were reweighted to match the pileup distribution observed in the data.

All MC samples underwent ATLAS detector simulation~\cite{SOFT-2010-01}
based on \textsc{Geant4}~\cite{Agostinelli:2002hh}.
The SUSY signal samples employed a fast simulation that parameterizes the response of the calorimeter~\cite{ATL-PHYS-PUB-2010-013}; the SM background samples used full \textsc{Geant4}
simulation.
\textsc{EvtGen}~v1.2.0~\cite{Lange:2001uf} was employed to model the decay of bottom and charm hadrons
in all samples except those generated by \SHERPA, which uses its internal modeling.
%-------------------------------------------------------------------------------

%-------------------------------------------------------------------------------
\section{Event reconstruction}
\label{sec:selection}
Candidate events are required to have at least one $pp$ interaction vertex
reconstructed with a minimum of two associated tracks with $\pt > 400$ MeV. The vertex with the highest $\sum\pt^2$ of the associated
tracks is selected as the primary vertex of the event.

This analysis defines two categories of identified leptons and jets, referred to as
\emph{preselected} and \emph{signal}, where signal leptons and jets are a subset of
preselected leptons and jets, respectively.

Preselected electrons are reconstructed with $\pt >
4.5~\GeV$ and within the pseudorapidity range of $|\eta| < 2.47$.
Furthermore, they are required to pass the likelihood-based \emph{VeryLoose}
identification, which is similar to the likelihood-based \emph{Loose} identification
defined in Ref.~\cite{Aaboud:2016vfy} but has a higher electron identification
efficiency. The likelihood-based electron identification criteria are based on calorimeter
shower shape and inner detector track information.
Preselected muons are identified using the \emph{Medium} criterion defined in
Ref.~\cite{PERF-2015-10} and required to satisfy $\pt > 4~\GeV$ and $|\eta|<
2.5$. The longitudinal impact parameter $z_0$ relative to the primary vertex
must satisfy $|z_0 \sin \theta| < 0.5$~mm for both the electrons and muons.

Preselected jets are reconstructed from calorimeter topological clusters~\cite{PERF-2014-07} 
in the region $|\eta| < 4.5$ using the
anti-$k_t$ algorithm~\cite{Cacciari:2008gp,Cacciari:2011ma} with radius parameter $R = 0.4$.
The jets are required to have $\pt>20$~\GeV{} after being calibrated in accord with
Ref.~\cite{PERF-2016-04} and having the expected energy contribution from pileup subtracted 
according to the jet area~\cite{PERF-2014-03}. In order to
suppress jets due to pileup, jets with $\pt < 60~\GeV$ and $|\eta| < 2.4$ are required to satisfy the \emph{Medium} working point of the jet vertex
tagger~\cite{PERF-2014-03}, which uses information from the tracks
associated to the jet. 
To reject events with detector noise or non-collision backgrounds,
events are rejected if they fail basic quality criteria~\cite{DAPR-2012-01}.

Jets that contain a $b$-hadron, referred to as
$b$-jets, are identified within $|\eta| < 2.5$ using the \emph{MV2c10}
algorithm~\cite{PERF-2012-04,ATL-PHYS-PUB-2016-012}. The working point is chosen so that $b$-jets from
simulated $t\bar{t}$ events are identified with an 85\% efficiency, with rejection factors of 3 for
charm-quark jets and 34 for light-quark and gluon jets.

The following procedure is used to resolve ambiguities between the
reconstructed leptons and jets. It employs the distance measure 
$\Delta R_y = \sqrt{(\Delta y)^2 + (\Delta \phi)^2}$,
where $y$ is the rapidity.
Electrons that share an inner detector track with a muon candidate
are discarded to remove bremsstrahlung from muons followed by a photon
conversion into electron pairs. Non-$b$-tagged jets that are separated from the remaining electrons
by $\Delta R_y < 0.2$ are removed. Jets that lie $\Delta R_y < 0.4$ from a muon
candidate and contain fewer than three tracks with $\pt>500$~MeV are removed to suppress
muon bremsstrahlung. Electrons or muons that lie $\Delta R_y < 0.4$ from surviving jet candidates are removed
to suppress bottom and charm hadron decays. 

Additional requirements on leptons that survive preselection are optimized for signal efficiency and background rejection.
Signal electrons must satisfy the \emph{Tight} identification criterion~\cite{ATL-PHYS-PUB-2016-015},
and be compatible with
originating from the primary vertex, with the significance 
of the transverse impact parameter defined relative to the beam position  
satisfying \mbox{$|d_0|/\sigma(d_0) < 5$}. 
From the remaining preselected muons, signal muons must satisfy
\mbox{$|d_0|/\sigma(d_0) < 3$}. 

The \emph{GradientLoose} and \emph{FixedCutTightTrackOnly} isolation criteria, as
detailed in Ref.~\cite{PERF-2015-10}, are imposed on signal electrons
and muons, respectively, to reduce contributions from fake/nonprompt leptons arising from jets 
misidentified as leptons, photon conversions, or semileptonic decays of heavy-flavor hadrons. 
These isolation requirements are either based on the presence of
additional tracks or based on clusters of calorimeter energy depositions inside a small cone around the
lepton candidate. Contributions from any other preselected leptons are
excluded in order to preserve efficiencies for signals with small dilepton
invariant mass.

After all lepton selection criteria are applied, the efficiency for reconstructing and identifying
signal electrons within the detector acceptance in the Higgsino and slepton
signal samples range from 15\% for $\pt=4.5~\GeV$ to over 70\% for $\pt>30~\GeV$.  
The corresponding efficiency for signal muons ranges from approximately 50\% at
$\pt=4~\GeV$ to over 85\% for $\pt>30~\GeV$.
Of the total predicted background, the fraction due to fake/nonprompt electrons
in an event sample with opposite-sign, different-flavor leptons 
falls from approximately 80\% at $\pt=4.5~\GeV$ to less than 5\% for $\pt>30~\GeV$,
while the fraction of fake/nonprompt muons in the same sample falls from 80\% at $\pt=4~\GeV$
to less than 8\% for $\pt>30~\GeV$.

From the sample of preselected jets, signal jets are selected if they satisfy
$\pt > 30~\GeV$ and $|\eta| < 2.8$, 
except for $b$-tagged jets where the preselected jet requirement of $\pt > 20~\GeV$ is maintained to maximize the
rejection of the \ttbar background.

Small corrections are applied to reconstructed electrons, muons, and
$b$-tagged jets in the simulated samples to match the reconstruction
efficiencies in data. The corrections for $b$-tagged jets account for the
differences in $b$-jet identification efficiencies as well as mis-identification
rates of $c$-, and light-flavour / gluon initiated jets between data and
simulated samples. The corrections for low-momentum leptons are obtained from
$J/\psi\rightarrow ee/\mu\mu$ events with the same tag-and-probe methods as used for higher-\pt{}
electrons~\cite{Aaboud:2016vfy} and muons~\cite{PERF-2015-10}.

The missing transverse momentum \ptmiss, with magnitude \met, is defined as the negative vector sum of the transverse momenta of all
reconstructed objects (electrons, muons and jets) and an additional soft term. 
The soft term is constructed from all tracks that are not associated with any object, 
but that are associated with the primary vertex. 
In this way, \met\ is adjusted for the best calibration of the jets and the
other identified physics objects above, while maintaining pileup independence in the soft term~\mbox{\cite{ATL-PHYS-PUB-2015-027}}.

%--------------------------------------------
\section{Signal region selection}
%--------------------------------------------
\label{sec:SRs}

\begin{table}[]
\centering\small
\caption{Summary of event selection criteria. The binning scheme used to define the final signal regions is shown in Table~\ref{tab:SRMLLMT2}.  Signal leptons and signal jets are used when applying all requirements.}
\label{tab:2LSRselection}
\vspace{0.2cm}
\begin{tabular}{lll}
\toprule
Variable                          & Common requirement\\
\midrule
Number of leptons                 & $=2$\\
Lepton charge and flavor         & $e^+ e^-$ or $\mu^+ \mu^-$\\
Leading lepton $\pt^{\ell_1}$     & \multicolumn{2}{l}{$>5$ (5)~\GeV{} for electron (muon)}\\
Subleading lepton $\pt^{\ell_2}$  & \multicolumn{2}{l}{$>4.5$ (4)~\GeV{} for electron (muon)}\\
\DRll                             & $>0.05$\\
\mll                              & \multicolumn{2}{l}{$\in[1,60]$~GeV excluding $[3.0,3.2]$~\GeV}\\
\met                              & $>200$~\GeV\\
Number of jets                    & $\geq 1$\\
Leading jet $\pt$                 & $>100$~\GeV\\
$\dphijetmet$                     & $>2.0$\\
$\mindphijetsmet$                 & $>0.4$\\
Number of $b$-tagged jets         & $=0$\\
$m_{\tau\tau}$                    & $<0$ or $>160$~\GeV\\
\midrule
                       & Electroweakino SRs                 & Slepton SRs \\
\midrule
\DRll                  & $<2$                               & --- \\
\mtlone                & $<70$~\GeV                          & --- \\
\met/\Htlep            & $>\max\left(5,15-2\frac{\mll}{1~\GeV}\right)$ & $>\max\left(3,15-2\left(\frac{\mtt^{100}}{1~\GeV}-100\right)\right)$ \\
Binned in              & $m_{\ell\ell}$      & $\mtt^{100}$ \\
\bottomrule
\end{tabular}
\end{table}

Table~\ref{tab:2LSRselection} summarizes the event selection criteria for all signal regions (SRs).
A candidate event is required to contain exactly two preselected same-flavor opposite-charge leptons
($e^+ e^-$ or $\mu^+ \mu^-$),
both of which must also be signal leptons.
In the SUSY signals considered, the highest sensitivity from this selection arises from 
two leptons produced either by the $\chitz$ decay via an off-shell $Z$ boson, or by the
slepton decays. 
The lepton with the higher (lower) \pt of each pair 
is referred to as the leading (subleading) lepton and is denoted by $\ell_1$
($\ell_2$).
The leading lepton is required to have $\pt^{\ell_1} > 5$~\GeV, 
which suppresses background due to fake/nonprompt leptons.
The \pt\ threshold for the subleading lepton remains at 4.5~\GeV{} for electrons
and 4~\GeV{} for muons to retain signal acceptance.
Requiring the separation \DRll\ between the two leptons to be greater than 0.05
suppresses nearly collinear lepton pairs originating from photon conversions or
muons giving rise to spurious pairs of tracks with shared hits. The invariant
mass \mll\ of the lepton pair is required to be greater than 1~\GeV{} for the same reason.  The dilepton invariant mass is further required to be
outside of the $[3.0, 3.2]$~\GeV{} window to suppress contributions from
$J/\psi$ decays, and less than 60~\GeV{} to suppress contributions from on-shell
$Z$ boson decays.  No veto is implemented around other resonances such as $\Upsilon$ or $\psi$ states, which
are expected to contribute far less to the SRs.

The reconstructed \met\ is required to be greater than 200~\GeV, where the efficiency
of the triggers used in the analysis exceeds 95\%.
For signal events to pass this \met\ requirement, the two \chioz\ momenta must align
by recoiling against hadronic initial-state radiation.
This motivates the requirements on the leading jet (denoted by $j_1$) of
$\pt^{j_1} > 100$~\GeV{} and $\dphijetmet>2.0$, where $\dphijetmet$ is the
azimuthal separation between $j_1$ and \ptmiss.
In addition, a minimum azimuthal separation requirement $\mindphijetsmet>0.4$
between any signal jet in the event and \ptmiss{} reduces the effect of
jet-energy mismeasurement on \met.

The leading sources of irreducible background are \ttbar, single-top,
$WW$/$WZ$ + jets (hereafter referred to as $WW$/$WZ$), and
$Z^{(*)}/\gamma^*(\to\tau\tau)$ + jets.
The dominant source of reducible background arises from processes where one or more leptons 
are \mbox{fake/nonprompt}, such as in $W$ + jets production.

Events containing $b$-tagged jets are rejected to reduce the \ttbar{} and
single-top background.
The \mbox{$Z^{(*)}/\gamma^*(\to\tau\tau)$} + jets background is suppressed
using the \mtautau{} variable~\cite{Han:2014kaa,Baer:2014kya,Barr:2015eva}, defined by
\mbox{$m_{\tau\tau} = \text{sign}\left(m_{\tau\tau}^2\right)\sqrt{\left|m_{\tau\tau}^2\right|}$},
which is the signed square root of $m_{\tau\tau}^2 \equiv 2p_{\ell_1} \cdot p_{\ell_2}(1+\xi_1)(1+\xi_2)$,
where $p_{\ell_1}$ and $p_{\ell_2}$ are the lepton four-momenta, 
while the parameters $\xi_1$ and $\xi_2$ are determined by solving
$\mathbf{p}_\text{T}^\textrm{miss} = \xi_1 \mathbf{p}_\textrm{T}^{\ell_1} + \xi_2 \mathbf{p}_\mathrm{T}^{\ell_2}$.
The definition of \mtautau{} approximates the invariant mass of a leptonically decaying $\tau$-lepton pair
if both $\tau$-leptons are sufficiently boosted so that the daughter neutrinos from each $\tau$
decay are collinear with the visible lepton momentum.
The \mtautau{} variable can take negative values in events where one of the lepton momenta
has a smaller magnitude than $\met$ and points in the hemisphere opposite to the \ptmiss{} vector.
Events with $0<\mtautau<160$~\GeV{} are rejected.
After the common and electroweakino SR selections in Table~\ref{tab:2LSRselection} are applied,
this veto retains 75\% of the Higgsino signal with $m(\chitz) = 110~\GeV$ and $m(\chioz) = 100~\GeV$, 
while 87\% of the $Z^{(*)}/\gamma^*(\to\tau\tau)$ + jets background is rejected.

After applying the common selection requirements above, two sets of SRs are
constructed to separately target the production of electroweakinos and sleptons.

In electroweakino production, the two leptons originating from $Z^*\to\ell\ell$ are both
soft, and their invariant mass is small. Due to the recoil of the SUSY particle
system against a jet from initial-state radiation,
the angular separation \DRll\ between the two leptons is required to be smaller
than 2.0.
The transverse mass of the leading lepton and \met, defined as $\mtlone =
\sqrt{2(E_\mathrm{T}^{\ell_1}\met -
\mathbf{p}_\mathrm{T}^{\ell_1}\cdot\ptmiss)}$, is required to be smaller than
70~\GeV{} to reduce the background from \ttbar, $WW/WZ$, and $W$ + jets.
The dilepton invariant mass \mll\ is correlated with $\Delta m(\chitz,\chioz)$, 
illustrated in Figure~\ref{fig:samples:invMas}, and is used to define the binning of the electroweakino SRs as
further described below.

\begin{table}[t!]
\centering\small
\caption{\label{tab:SRMLLMT2}
Signal region binning for the electroweakino and slepton SRs.
Each SR is defined by the lepton flavor ($ee$, $\mu\mu$, or $\ell\ell$ for both) and a range of
\mll\ (for electroweakino SRs) or \mtth\ (for slepton SRs) in \GeV.
The inclusive bins are used to set model-independent limits, 
while the exclusive bins are used to derive exclusion limits on signal models.  }
\vspace{0.2cm}
\resizebox{\linewidth}{!}{
\renewcommand{\arraystretch}{1.2}
\begin{tabular}{llccccccc}
\toprule                  
\multicolumn{8}{c}{Electroweakino SRs}                                \\
\midrule
Exclusive & \SReemll, \SRmmmll & $[1,3]$ & $[3.2,5]$ & $[5,10]$ & $[10,20]$ & $[20,30]$ & $[30,40]$ & $[40,60]$ \\
Inclusive & \SRllmll & $[1,3]$ & $[1,5]$   & $[1,10]$ & $[1,20]$  & $[1,30]$  & $[1,40]$ & $[1,60]$ \\
\midrule
\multicolumn{8}{c}{Slepton SRs}                           \\
\midrule
Exclusive & \SReemtt, \SRmmmtt & & $[100,102]$ & $[102,105]$ & $[105,110]$ & $[110,120]$ &
$[120,130]$ & $[130,\infty]$ \\
Inclusive & \SRllmtt & &  $[100,102]$ & $[100,105]$ & $[100,110]$ & $[100,120]$
& $[100,130]$ & $[100,\infty]$ \\
\bottomrule
\end{tabular}
}
\end{table}

In slepton pair production,
the event topology can be
used to infer the slepton mass given the LSP mass.
The stransverse mass~\cite{Lester:1999tx, Barr:2003rg} is defined by
\newcommand{\ptvec}{\mathbf{p}_\mathrm{T}}
\newcommand{\qtvec}{\mathbf{q}_\mathrm{T}}
\begin{linenomath*}
\begin{equation*}
   \mtt^{m_\chi} \left(\ptvec^{\ell_1},\ptvec^{\ell_2},\ptvec^\mathrm{miss}\right)
   = \underset{\qtvec}{\min}
		\left(\max\left[
   			\mt\left(\ptvec^{\ell_1},\qtvec,m_\chi\right),
   			\mt\left(\ptvec^{\ell_2},\ptmiss-\qtvec,m_\chi\right)
   		\right]\right),
\end{equation*}
\end{linenomath*}
where $m_\chi$ is the hypothesized mass of the invisible particles,
and the transverse vector $\qtvec$ with magnitude $q_\text{T}$ is chosen to minimize the larger of the two transverse masses, defined by
\begin{linenomath*}
\begin{equation*}
  \mt\left(\ptvec,\qtvec,m_\chi\right)
  = \sqrt{ m_\ell^2 + m_\chi^2
           + 2\left(\sqrt{\pt^2 + m_\ell^2} \sqrt{q_\mathrm{T}^2 + m_\chi^2} -
                    \ptvec \cdot \qtvec
              \right)
         }.
\end{equation*}
\end{linenomath*}
For events arising from signals with slepton mass $m(\slepton)$ and LSP mass $m(\chioz)$, 
the values of $m_\text{T2}^{m_\chi}$ are bounded from above by $m(\slepton)$ 
when $m_\chi$ is equal to $m(\chioz)$, 
i.e.\  $m_\text{T2}^{m_\chi} \leq m(\slepton)$ for $m_\chi = m(\chioz)$. 
The stransverse mass with $m_\chi = 100$~\GeV, denoted $\mtt^{100}$, is used 
to define the binning of the slepton SRs as further described below. 
The chosen value of 100~\GeV{} is based on the expected LSP masses of the signals targeted by this analysis. 
The distribution of $\mtt^{100}$ does not vary significantly for
signals where $m(\chioz) \neq 100~\GeV$.

\begin{figure}[tbp]
  \centering
  \begin{subfigure}{0.49\textwidth}
    \includegraphics[width=\textwidth]{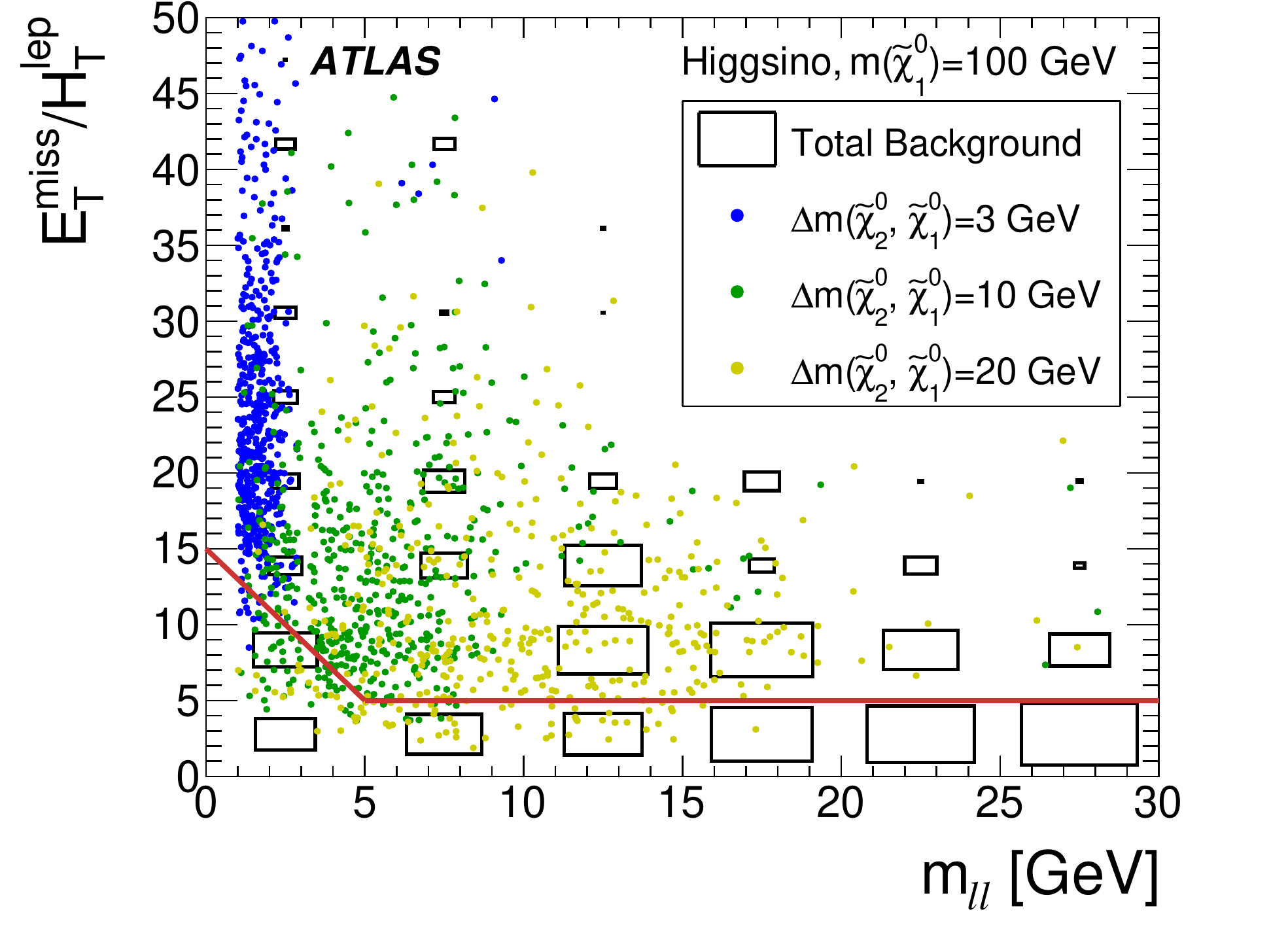}
  \end{subfigure}
  \begin{subfigure}{0.49\textwidth}
    \includegraphics[width=\textwidth]{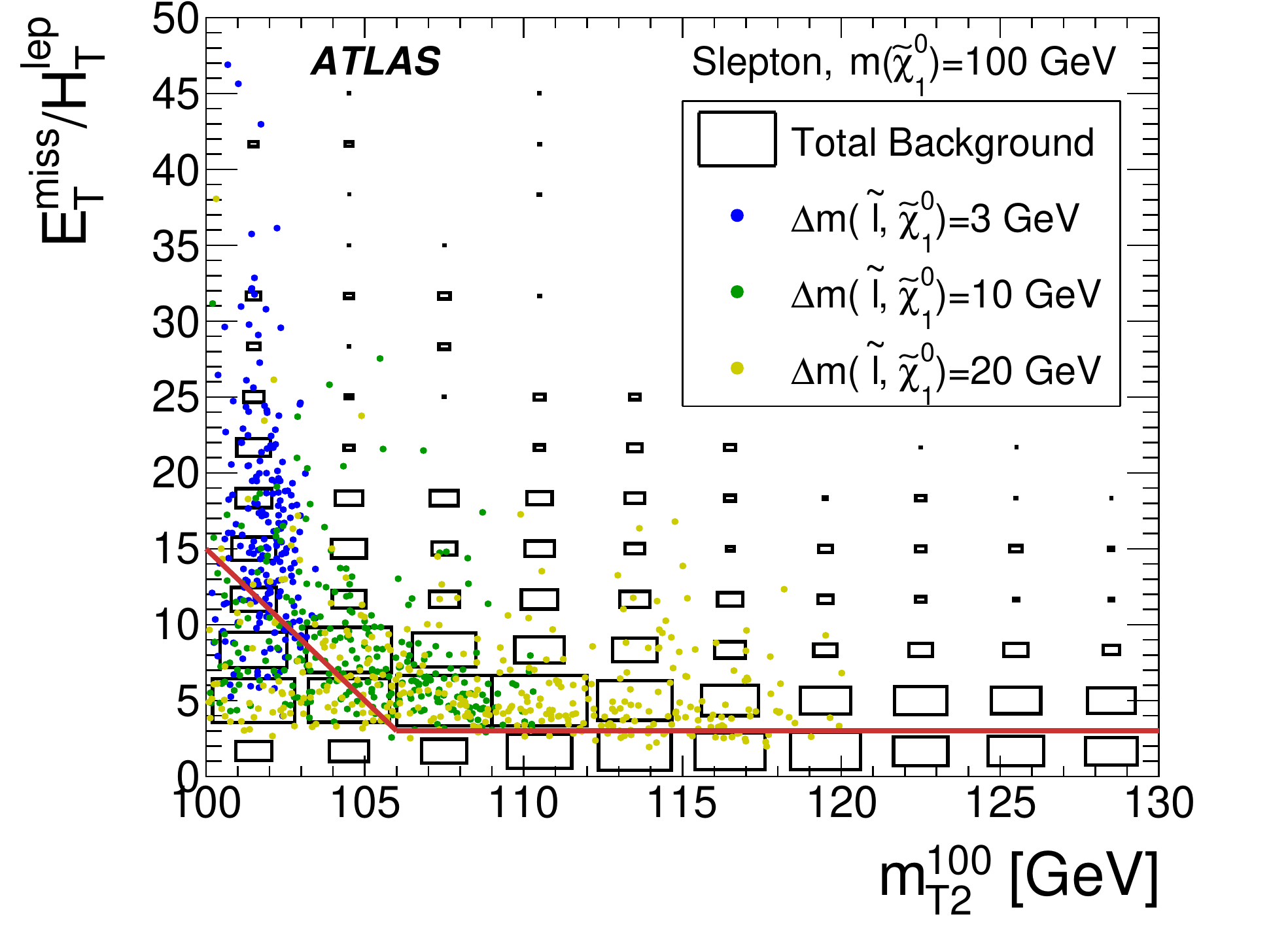}
  \end{subfigure}
  \caption{Distributions of $\met/\htlep$ for the electroweakino (left) and
  slepton (right) SRs, after applying all signal region selection criteria except those on
  $\met/\htlep$, $\mll$, and $\mttwo$. The solid red line indicates the requirement
  applied in the signal region; events in the region below the red line are rejected.
  Representative benchmark signals for the Higgsino (left) and slepton (right) simplified models
  are shown as circles.
  Both signal and background are normalized to their expected yields in 36.1~\ifb.
  The total background includes the MC
  prediction for all the processes listed in Table~\ref{tab:MCconfig} and a
  data-driven estimate for fake/nonprompt leptons discussed further in
  Section~\ref{sec:background}.    }
  \label{fig:METoverHTLep2D}
\end{figure}

The scalar sum of the lepton transverse momenta $\Htlep = \pt^{\ell_1} + \pt^{\ell_2}$ is
smaller in compressed-scenario SUSY signal events than in background events such as
SM production of $WW$ or $WZ$.
The ratio $\met/\Htlep$ provides signal-to-background discrimination which
improves for smaller mass splittings in the signals and is therefore used as a
sensitive variable in both the electroweakino and slepton SRs.
The minimum value of the $\met/\Htlep$ requirement is adjusted event by event according to the
size of the mass splitting inferred from the event kinematics.
For the electroweakino SRs, this is achieved with \mll\ as $\met/\Htlep >
\max[5,15-2\mll/(1~\GeV)]$.
For the slepton SRs, $\mtth-100$~\GeV{} is used as $\met/\Htlep >
\max[3,15-2\{\mtth/(1~\GeV)-100\}]$.
Figure~\ref{fig:METoverHTLep2D} illustrates the $\met/\htlep$ requirement for
electroweakino and slepton SRs.

Table~\ref{tab:SRMLLMT2} defines the binning of the SRs.
The electroweakino SRs are divided into seven non-overlapping ranges of \mll,
which are further divided by lepton flavor ($ee$, $\mu\mu$), and referred to as exclusive regions.
Seven inclusive regions are also defined, characterized by overlapping ranges of \mll.
For the slepton SRs, $\mtth$ is used to define 12 exclusive regions and 6 inclusive regions.
When setting model-dependent limits on the electroweakino (slepton) signals, 
only the exclusive \SReemll\ and \SRmmmll\ regions (\SReemtt\ and \SRmmmtt\ regions)
are statistically combined in a simultaneous fit.
When setting model-independent upper limits on new physics signals,
only the inclusive \SRllmll\ and \SRllmtt\ regions are considered.
The details of these statistical procedures are given in Section~\ref{sec:results}.

After all selection criteria are applied, the Higgsino model
with $m(\chitz)=110~\GeV$ and $m(\chioz)=100~\GeV$ 
has an acceptance times efficiency of $6.5\times10^{-5}$ in \SRllmll\ $[1,60]$. 
The acceptance times efficiency in \SRllmtt\ $[100,\infty]$ for the slepton model, with $m(\tilde{\ell})=110~\GeV$ and
$m(\chioz)=100~\GeV$, is $3.3\times10^{-3}$.  
%-------------------------------------------------------------------------------

%-------------------------------------------------------------------------------
\section{Background estimation}
\label{sec:background}
A common strategy is used to determine the SM background in all SRs.
The dominant sources of irreducible background events that contain two prompt
leptons, missing transverse momentum and jets are $t\bar{t}$, $tW$, $WW/WZ$, and
$Z^{(*)}/\gamma^*(\to\tau\tau)$ + jets, which are estimated using MC
simulation. The main reducible backgrounds are from events containing fake/nonprompt
leptons. These processes are estimated collectively with a
data-driven method. While the fake/nonprompt lepton background tends to be
dominant at low values of \mll{} and \mtth, the irreducible $t\bar{t}$, $tW$, $WW/WZ$ processes are more important at the upper end of the distributions.

%--------------------------------------------
\subsection{Irreducible background}
%--------------------------------------------
The MC simulations of $t\bar{t}, tW$ and $Z^{(*)}/\gamma^*(\to\tau\tau)$ + jets
background processes are normalized in a simultaneous fit to the observed data counts in control regions (CRs) using statistical procedures detailed in Section~\ref{sec:results}. 
The CRs are designed to be statistically disjoint from the SRs, to be enriched in a particular background process,
to have minimal contamination from the signals considered, and to exhibit kinematic properties similar to the SRs.
The event rates in the SRs are then predicted by extrapolating from the CRs using the simulated MC distributions. 
This extrapolation is validated using events in dedicated validation regions (VRs),
which are not used to constrain the fit and are orthogonal in selection to the
CRs and SRs. The definitions of these regions are summarized in Table~\ref{tab:CRs}.

The $t\bar{t}$ and $tW$, diboson $WW/WZ$, and
$Z^{(*)}/\gamma^*(\to\tau\tau)$ + jets processes containing two prompt leptons 
all yield same-flavor lepton
pairs ($ee$ and $\mu\mu$) at the same rate as for different-flavor pairs ($e\mu$ and $\mu e$, where the first lepton is the leading lepton).
To enhance the statistical constraining power of the respective CRs, 
all possible flavor assignments ($ee$, $\mu\mu$, $e\mu$, and $\mu e$)  are selected when defining the CRs.

\begin{figure}[t!]
 \begin{center}
  \includegraphics[width=0.49\textwidth]{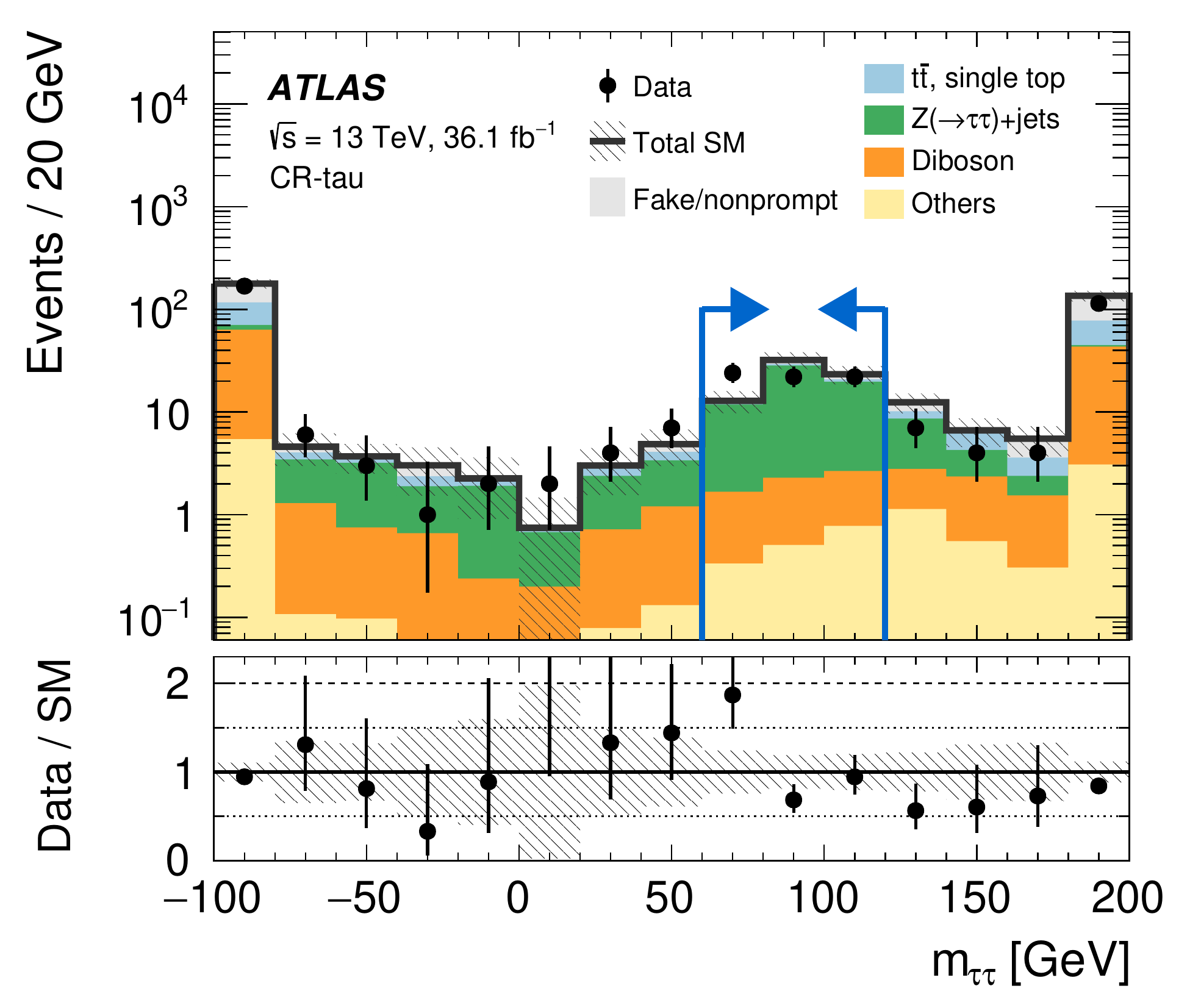}  
  \includegraphics[width=0.49\textwidth]{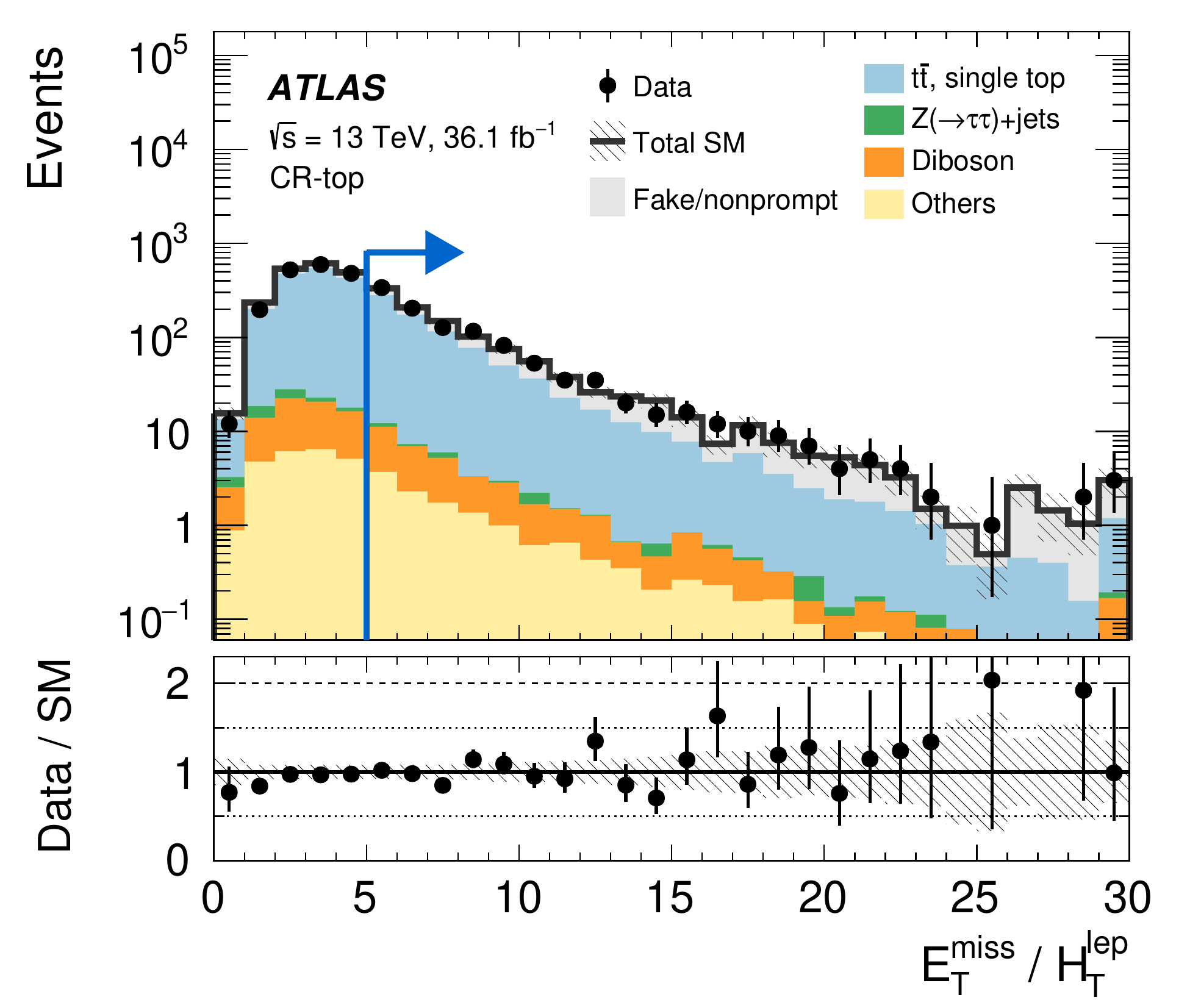}    
  \end{center}
\caption{Examples of kinematic distributions after the background-only fit 
showing the data as well as the expected
background in control regions CR-tau (left) and CR-top (right). The
full event selection of the corresponding regions is applied, except for the
requirement that is imposed on the variable being
plotted. This requirement is indicated by blue arrows in the distributions. The
first (last) bin includes underflow (overflow).
Background processes containing fewer than two prompt leptons are categorized as `Fake/nonprompt'.
The category
`Others' contains rare backgrounds from triboson, Higgs boson, and the remaining top-quark 
production processes listed in Table~\ref{tab:MCconfig}. The uncertainty
bands plotted include all statistical and systematic uncertainties. 
}
 \label{sec:results:distributionsCRs}
\end{figure}

Two single-bin CRs are considered, which have all the selections
in Table~\ref{tab:2LSRselection} applied 
unless stated otherwise in Table~\ref{tab:CRs}.
A sample enriched in top quarks with 71\% purity, CR-top, is defined by selecting events with at least one $b$-tagged jet.
This CR has 1100 observed events and is used to constrain the normalization of the \ttbar\ and $tW$ processes
with dilepton final states. 
A sample enriched in the $Z^{(*)}/\gamma^*(\to\tau\tau)$ + jets processes with
83\% purity, CR-tau, is constructed by selecting events satisfying $60 < m_{\tau\tau} < 120$~\GeV. 
This CR has 68 observed events and the variable $\met/\Htlep$ is required to have a value between 4 and 8
to reduce potential contamination from signal events.
Figure~\ref{sec:results:distributionsCRs} shows the background
composition of the CR-tau and CR-top regions. The signal contamination in both regions is
typically below 3\% and is at most 11\%.

\begin{table}[]
\centering\small
\caption{\label{tab:CRs}
Definition of control and validation regions.
The common selection criteria in Table~\ref{tab:2LSRselection} are implied unless otherwise specified.}
\vspace{0.2cm}
\begin{tabular}{llll}
\toprule
Region     & Leptons
           & $\met/\Htlep$ & Additional requirements \\
\midrule
CR-top     & $e^\pm e^\mp$, $\mu^\pm \mu^\mp$, $e^\pm \mu^\mp$, $\mu^\pm e^\mp$
           & $>5$       & $\ge 1$ $b$-tagged jet(s) \\
CR-tau     & $e^\pm e^\mp$, $\mu^\pm \mu^\mp$, $e^\pm \mu^\mp$, $\mu^\pm e^\mp$
           & $\in[4,8]$ & $\mtautau \in[60,120]$~\GeV \\
\midrule
VR-VV      & $e^\pm e^\mp$, $\mu^\pm \mu^\mp$, $e^\pm \mu^\mp$, $\mu^\pm e^\mp$
           & $<3$       & \\
VR-SS      & $e^\pm e^\pm$, $\mu^\pm \mu^\pm$, $e^\pm\mu^\pm$, $\mu^\pm e^\pm$
           & $>5$       & \\
VRDF-\mll  & $e^\pm \mu^\mp$, $\mu^\pm e^\mp$
           & $>\max\left(5,15-2\frac{\mll}{1~\GeV}\right)$ & $\DRll<2$, $\mtlone<70$~\GeV \\
VRDF-\mtth & $e^\pm \mu^\mp$, $\mu^\pm e^\mp$
           & $>\max\left(3,15-2\left(\frac{\mtt^{100}}{1~\GeV}-100\right)\right)$ & \\
\bottomrule
\end{tabular}
\end{table}

It is difficult to select a sample of diboson events pure enough to be
used to constrain their contribution to the SRs.
The diboson background is therefore estimated with MC simulation.
A diboson VR, denoted by VR-VV, is constructed by requiring $\met/\Htlep < 3.0$.
This sample consists of approximately 40\% diboson events, 20\% fake/nonprompt
lepton events, 25\% $t\bar{t}$ and single-top events, 
and smaller contributions from $Z^{(*)}/\gamma^*\to\tau\tau$ and other processes. 
The signal contamination in VR-VV is at most 9\%.
This region is used to test the modeling of the diboson background
and the associated systematic uncertainties.

Additional VRs are constructed from events with different-flavor ($e\mu$ and $\mu e$) leptons.
These VRs, \mbox{VRDF-$\mll$} and \mbox{VRDF-\mtth}, are defined using the same selection criteria as the
electroweakino and slepton SRs, respectively, 
and are used to validate the extrapolation of background in the fitting procedure
within the same kinematic regime as the SRs. 
The electroweakino signal contamination in \mbox{VRDF-$\mll$} is always below 8\%,
while the slepton signal contamination in \mbox{VRDF-\mtth} is always negligible.

For each VR, the level of agreement between the kinematic distributions 
of data and predicted events is checked. 
The \mbox{VRDF-$\mll$} and \mbox{VRDF-\mtth} regions 
are also presented binned in $m_{\ell\ell}$ and $\mtt^{100}$, respectively,
using the same intervals as the exclusive SRs in Table~\ref{tab:SRMLLMT2} 
to ensure that these VRs consist of events with the same kinematic selection as
the SRs. 

%--------------------------------------------
\subsection{Reducible background}
%--------------------------------------------

Two sources of reducible background are considered: processes where
fake/nonprompt leptons are amongst the two selected signal leptons, 
and those where the reconstructed \met values are instrumental in origin.

The fake/nonprompt lepton background arises from jets misidentified as
leptons, photon conversions, or semileptonic decays of heavy-flavor hadrons.
Studies based on simulated samples indicate that the last of these is the dominant
component in the SRs. Since MC simulation is not expected to model these
processes accurately, the data-driven Fake Factor method~\cite{HIGG-2013-13} is employed.

The Fake Factor procedure first defines a tight set of criteria, labeled ID, corresponding to 
the requirements applied to signal leptons used in the analysis. 
Second, a loose set of criteria, labeled anti-ID, has one or more 
of the identification, isolation, or $|d_0|/\sigma(d_0)$ requirements
inverted relative to signal leptons to obtain an orthogonal sample enriched
in fake/nonprompt leptons.
The ratio of ID to anti-ID leptons defines the fake factor.

The fake factors are measured in events collected with prescaled single-lepton
triggers.
These single-lepton triggers have lepton identification requirements looser than those used in
the anti-ID lepton selection, and have \pt{} thresholds ranging from
4~\GeV{} to 20~\GeV.
This sample, referred to as the measurement region, is dominated by multijet events
with fake/nonprompt leptons.  Both the electron and muon fake factors are measured in this region
as a function of reconstructed lepton $\pt$.  
The muon fake factors are also found to have a dependence on the number of $b$-tagged jets in the event.
The fake factors used in CR-top are therefore computed in events with $>0~b$-tagged jets, while all other regions
use fake factors computed using events with zero $b$-tagged jets.

To obtain the fake/nonprompt lepton prediction in a particular region, these fake factors are applied to
events satisfying the corresponding selection requirements, except with an anti-ID lepton replacing an ID lepton.
MC studies indicate that the leptons in the anti-ID region arise from processes similar
to those for fake/nonprompt leptons passing the signal selection requirements
in the SR.
The contributions from prompt leptons that pass the ID and anti-ID requirements in the measurement region,
and that pass the anti-ID requirements in the region under study, are subtracted using MC simulation.
The yields from this procedure are cross-checked in VRs, named VR-SS,
which have similar kinematic selections as the SRs, but are enriched in fake/nonprompt leptons
by requiring two leptons with the same electric charge.
As the subleading lepton is found to be the fake/nonprompt lepton in most
cases, the VR-SS are divided into $ee+\mu e$ and $\mu\mu + e\mu$, where the left
(right) lepton of each pair denotes the leading (subleading) lepton. The
fraction of events in which both leptons are fake/nonprompt is found to be
small by considering the rate of anti-ID leptons in data.  The electroweakino signal contamination in VR-SS is
typically negligible, and always below 7\%, while the slepton signal contamination in VR-SS is always negligible.

Background processes with no invisible particles can satisfy the $\met>200$~\GeV{} requirement 
when the momenta of visible leptons or jets are mismeasured by the detector.
Contributions of these events in the SRs arising from processes
such as Drell--Yan dilepton production are studied with MC simulation and are found to be negligible.
This estimate is cross-checked with a data-driven method using independent
event samples defined by relaxed or inverted selection criteria.
A lower \met{} requirement is used to accept a higher rate of
$Z^{(*)}/\gamma^*\to\ell^+\ell^-$ events, while relaxed requirements on the
kinematics of the leading jet, \mtautau, $\MET/\HTlep$, and lepton isolation minimize the impact of any signal contamination.  The results from the data-driven method are consistent
with the estimates based on MC simulation.
%-------------------------------------------------------------------------------

%-------------------------------------------------------------------------------
\section{Systematic uncertainties}
\label{sec:systematics}
The sources of systematic uncertainty affecting the background and signal
predictions consist of uncertainties due to experimental sources, which include those
from the Fake Factor method, and uncertainties arising from the theoretical
modeling in simulated samples. 

The largest sources of experimental systematic uncertainty is the
fake/nonprompt background prediction from the Fake Factor procedure.
In this method, systematic uncertainties arise from the size of the samples used
to measure the fake factors, which are uncorrelated between events with respect to
the $\pt$ and flavor of the anti-ID lepton, but otherwise correlated across the
different CRs and SRs. Additional uncertainties are
assigned to account for differences in the event and lepton kinematics between the measurement region and SRs.
The differences between the Fake Factor prediction and observed data in the
VR-SS regions are used to assign additional systematic uncertainties. These
uncertainties are considered correlated across the different SRs, but
uncorrelated as regards the flavor of the anti-ID lepton in the event.  Uncertainties originating from the MC-based subtraction of prompt leptons in the Fake Factor measurement region are found to be negligible.

Further significant experimental systematic uncertainties are related to the jet
energy scale (JES) and resolution (JER), flavor-tagging, and the reweighting procedure
applied to simulated events to match pileup conditions observed in data.
Uncertainties in the lepton reconstruction and identification efficiencies,
together with energy/momentum scale and resolution also contribute, but are found to be small.  The systematic
uncertainties for low-momentum leptons are derived using the same procedure as
for higher-\pt{} electrons~\cite{Aaboud:2016vfy} and muons~\cite{PERF-2015-10}.

In addition to the experimental uncertainties, several sources of theoretical
modeling uncertainty affect the simulated samples of the
dominant SM backgrounds, i.e., \ttbar, $tW$, $Z^{(*)}/\gamma^*(\to\tau\tau)$ +
jets, and diboson processes.  The effects of the QCD renormalization and factorization scale
uncertainties are evaluated by independently varying the corresponding event generator 
parameters up and down by a factor of two. 
The impact of the uncertainty of the strong coupling
constant $\alphas$ on the acceptance is also considered. The effects of PDF uncertainties are
evaluated by reweighting the simulated samples to the CT14~\cite{Dulat:2015mca}
and MMHT2014~\cite{Harland-Lang:2014zoa} PDF sets and taking the envelope of the predicted yields.
The theoretical modeling uncertainties are evaluated in each of the CRs and SRs
and their effect is correlated for events across all regions.
For the dileptonic diboson background, the uncertainties of the normalization
and shape in the SRs are dominated by the QCD scale variations.
The normalization uncertainties of the top quark and
$Z^{(*)}/\gamma^*(\to\tau\tau)$ + jets contributions are constrained by the
simultaneous fit, and only the shape uncertainties relating the CRs to the SRs affect the results.

\begin{figure}[tbp]
  \centering
  \begin{subfigure}{0.49\textwidth}
    \includegraphics[width=\textwidth]{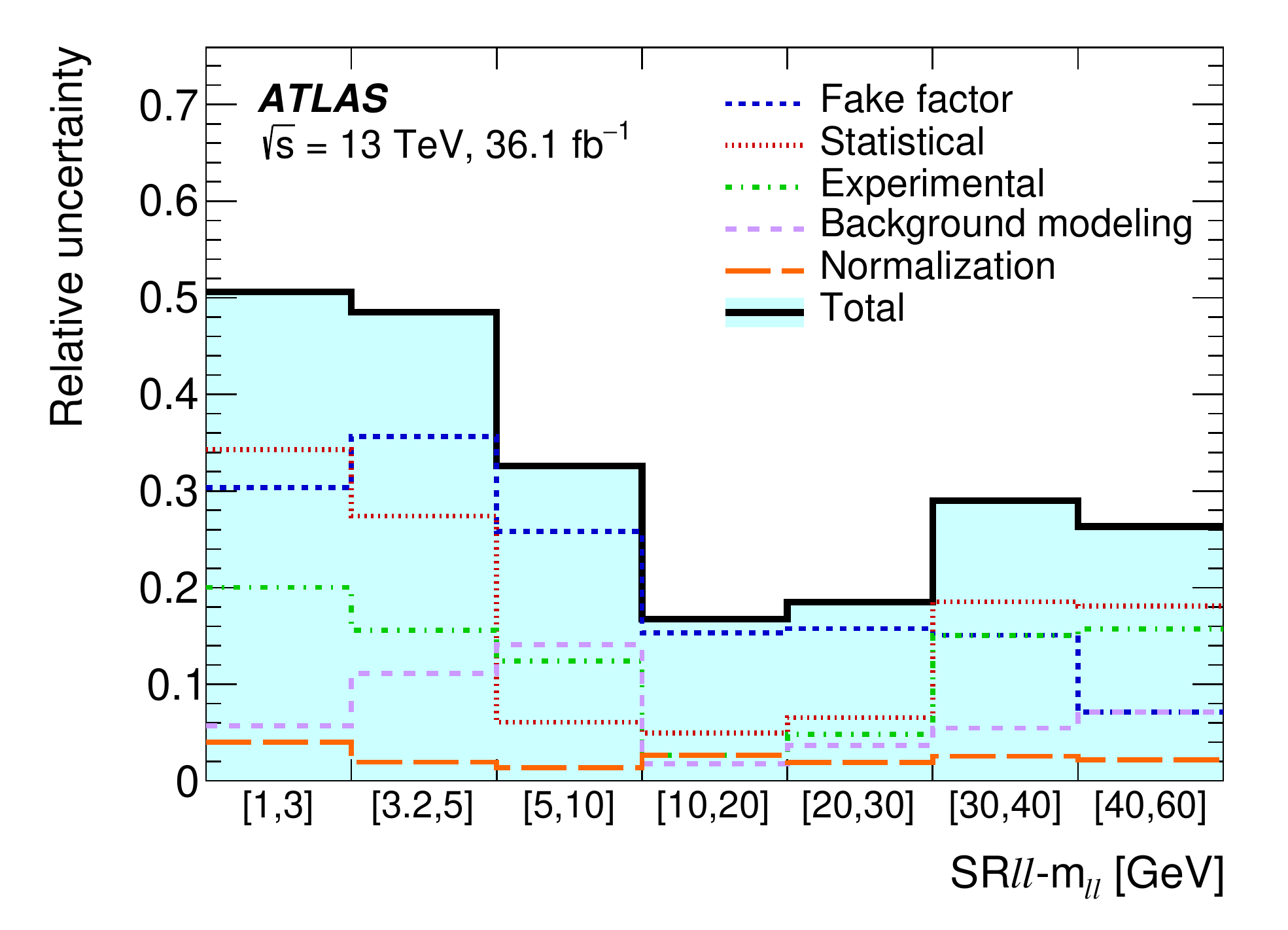}
  \end{subfigure}
  \begin{subfigure}{0.49\textwidth}
    \includegraphics[width=\textwidth]{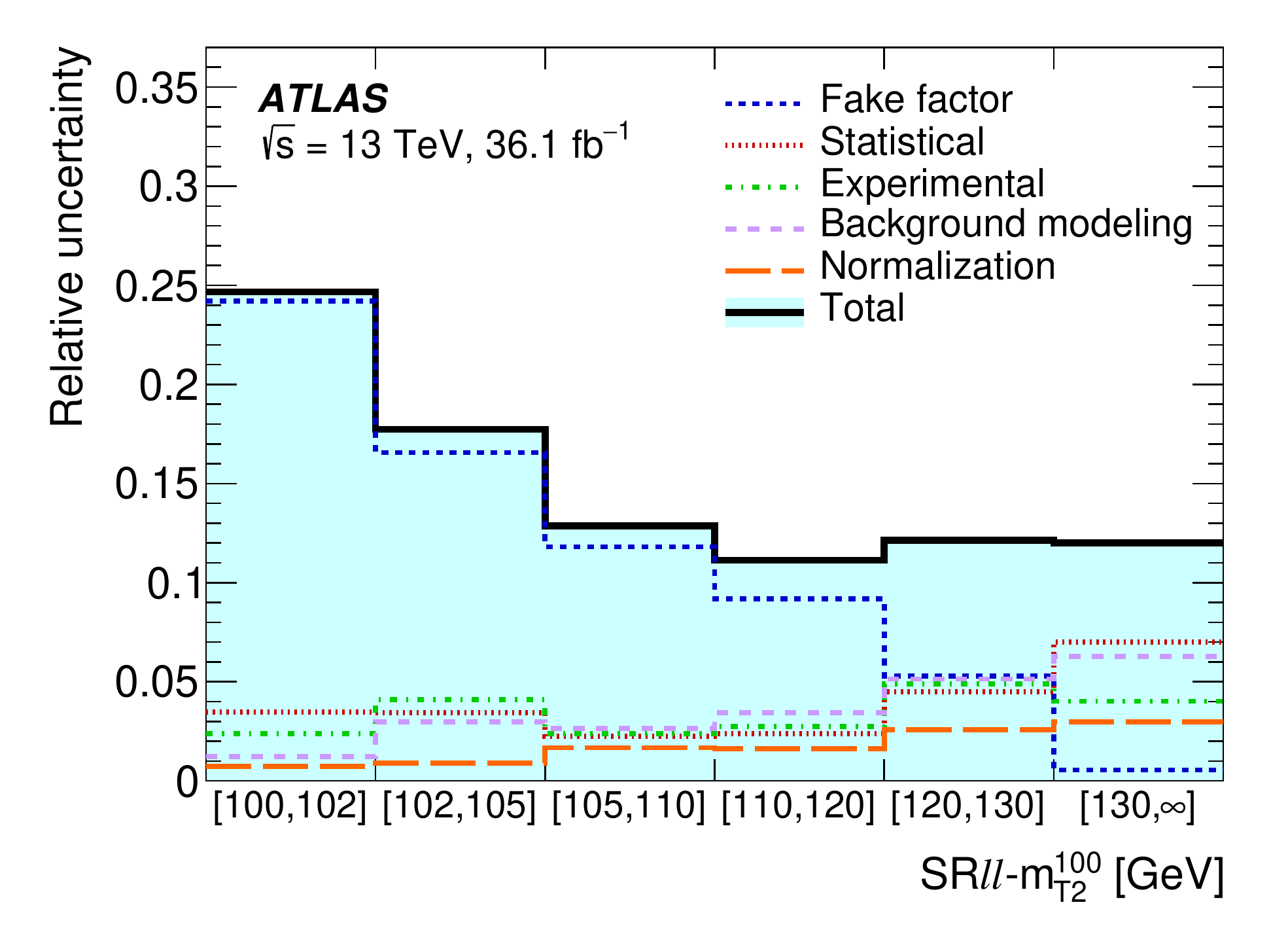}
  \end{subfigure}
  \caption{The relative systematic uncertainties in the
  background prediction in the exclusive electroweakino (left) and slepton
  (right) SRs.
  The individual uncertainties can be correlated and do not necessarily add up in
  quadrature to the total uncertainty. 
  }
  \label{fig:systematics}
\end{figure}

Figure~\ref{fig:systematics} shows the relative size of the various classes of
uncertainty in the background predictions in the exclusive
electroweakino and slepton SRs. The uncertainties related to the Fake
Factor method are displayed separately from the remaining experimental
uncertainties due to their relatively large contribution. The breakdown also
includes the uncertainties in the normalization factors of the
$Z^{(*)}/\gamma^*(\to\tau\tau)$ + jets and the combined $t\bar{t}$ and $tW$
backgrounds as obtained from CR-tau and CR-top, respectively. 

The theoretical modeling uncertainties in the expected
yields for SUSY signal models are estimated by varying by a factor of two
the {MG5\_aMC@NLO} parameters corresponding to the renormalization,
factorization and CKKW-L matching scales, as well as the \textsc{Pythia8}
shower tune parameters. The overall uncertainties in the signal acceptance 
range from about 20\% to 40\%
and depend on the SUSY particle mass splitting and the production process.
Uncertainties in the signal acceptance due to PDF uncertainties are evaluated
following the PDF4LHC15 recommendations~\cite{Butterworth:2015oua} and amount to
15\% at most for large \chitz~or $\widetilde{\ell}$ masses.
Uncertainties in the shape of the $\mll$ or $\mtt^{100}$ signal distributions due to the sources above are found
to be small, and are neglected.

%-------------------------------------------------------------------------------

\FloatBarrier

%-------------------------------------------------------------------------------
\section{Results and interpretation}
\label{sec:results}
The \textsc{HistFitter} package~\cite{Baak:2014wma} is used to implement the statistical interpretation based on a profile likelihood method~\cite{Cowan:2010js}.
Systematic uncertainties are treated as nuisance parameters in the likelihood.

To determine the SM background predictions independent of the SRs,
only the CRs are used to constrain the fit parameters by likelihood
maximization assuming no signal events in the CRs; this is referred to as the
background-only fit.
The normalizations, $\mu_{Z^{(*)}/\gamma^*\rightarrow\tau\tau}$ and
$\mu_\text{top}$, respectively for the $Z^{(*)}/\gamma^*(\to\tau\tau)$ + jets MC
sample and the combined $t\bar{t}$ and $tW$ MC samples are extracted in a simultaneous fit to the data events in CR-tau and CR-top, as
defined in Section~\ref{sec:background}.
The normalization parameters,
as obtained from the background-only fit, are
$\mu_{Z^{(*)}/\gamma^*\rightarrow\tau\tau}=0.72 \pm 0.13$ and
$\mu_{\mathrm{top}}=1.02\pm0.09$, whose uncertainties include statistical and systematic contributions combined.

\begin{figure}
    \centering
        \includegraphics[width=\textwidth]{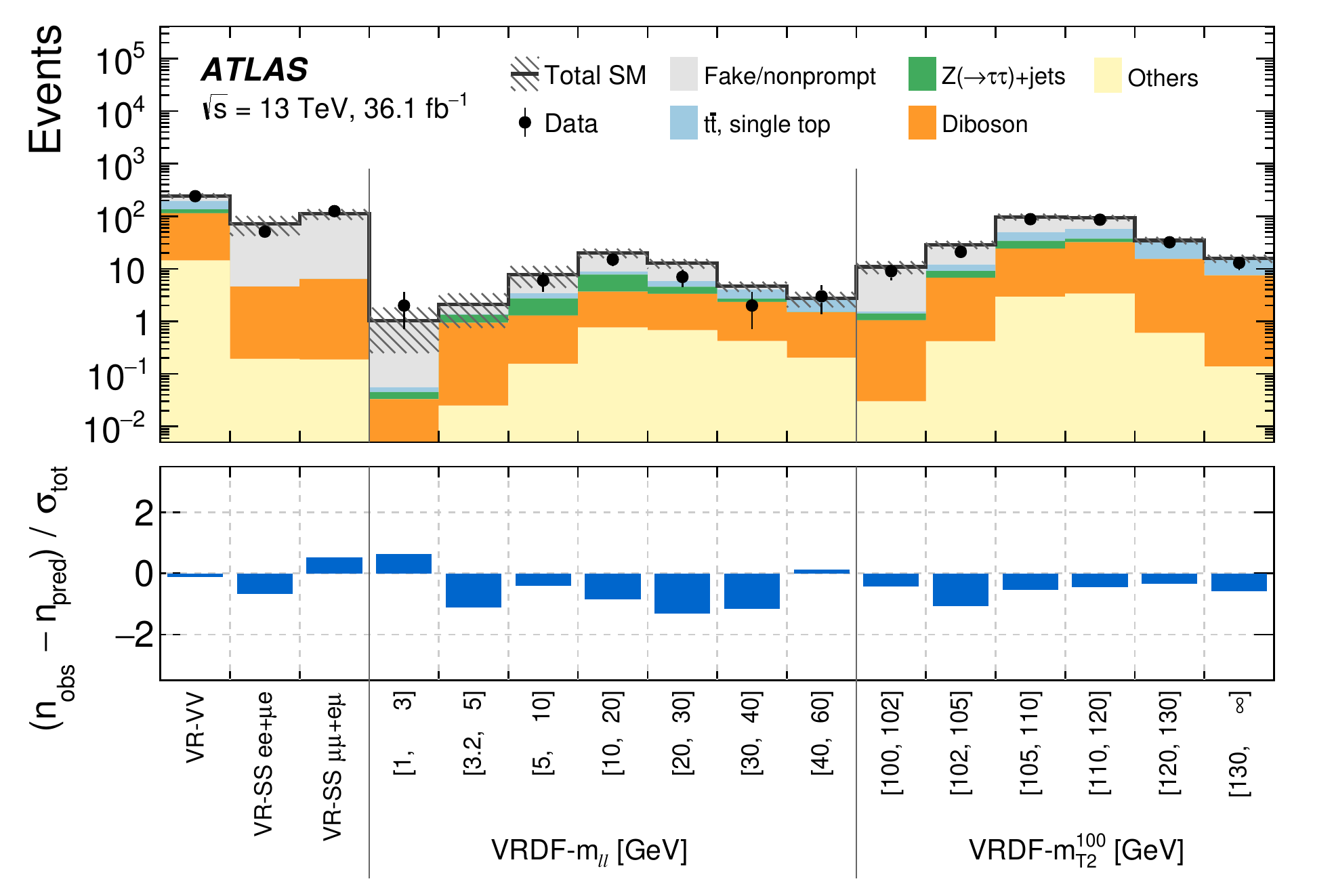}
        \caption{Comparison of observed and expected event yields in the
    validation regions after the background-only fit. 
    Background processes containing fewer than two prompt leptons are categorized as `Fake/nonprompt'.    
    The category
    `Others' contains rare backgrounds from triboson, Higgs boson, and the remaining top-quark 
    production processes listed in Table~\ref{tab:MCconfig}. Uncertainties in the
    background estimates include both the statistical and systematic uncertainties, 
    where $\sigma_\text{tot}$ denotes the total uncertainty.}
    \label{fig:pull_plot_summary_yields:vrs}
\end{figure}

\begin{figure}[h!]
 \begin{center}
  \includegraphics[width=0.49\textwidth]{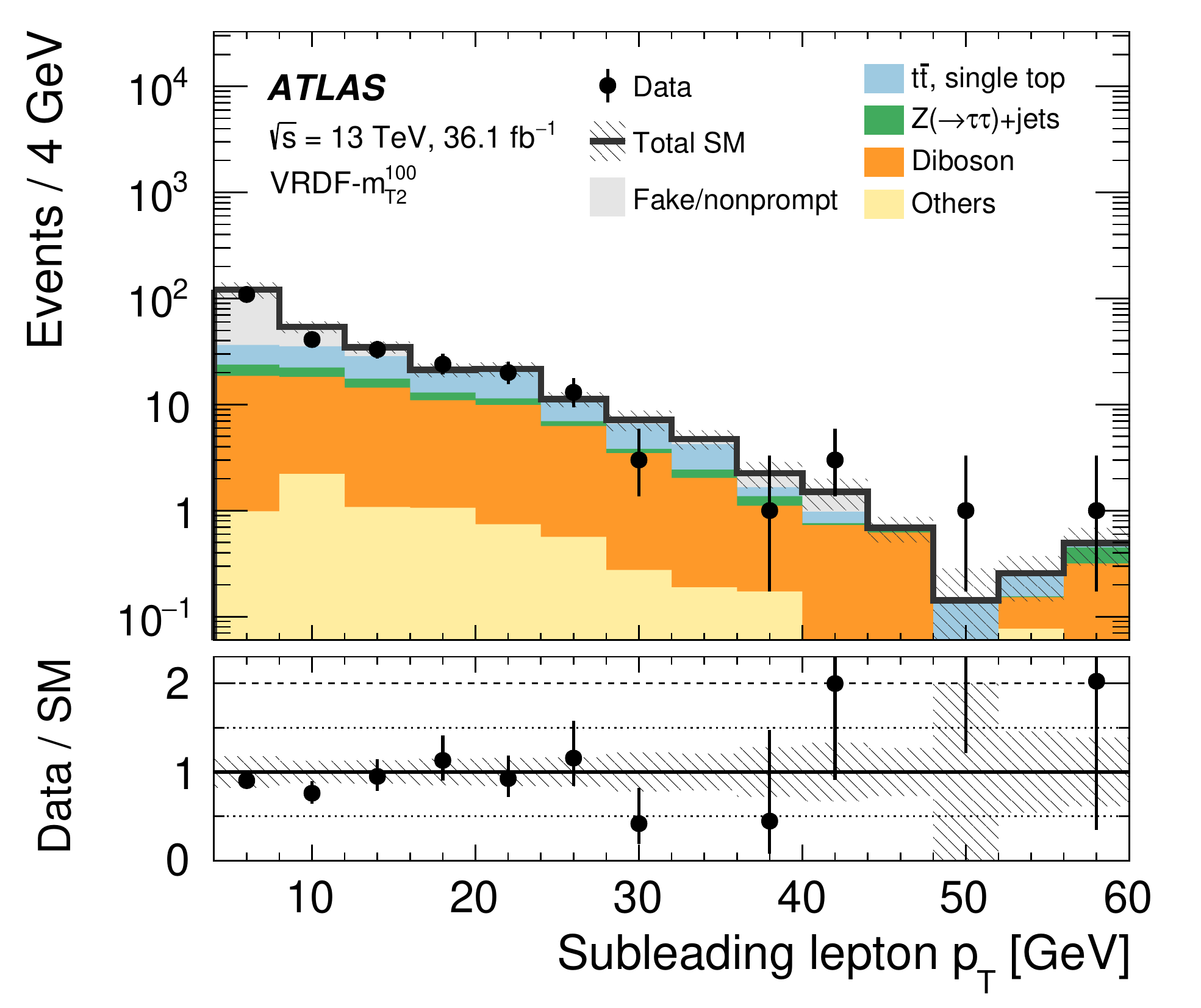}  
  \includegraphics[width=0.49\textwidth]{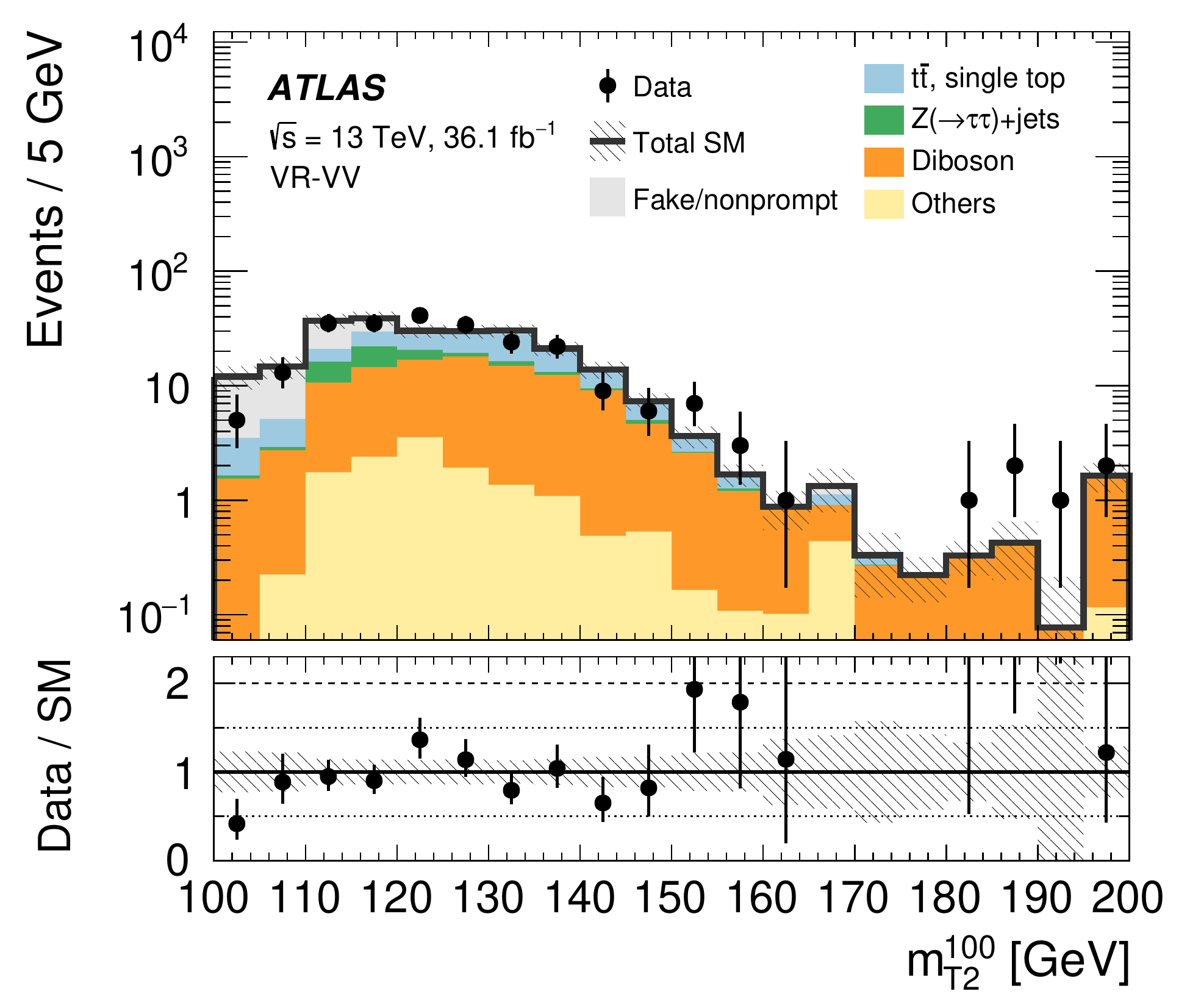}  
  \includegraphics[width=0.49\textwidth]{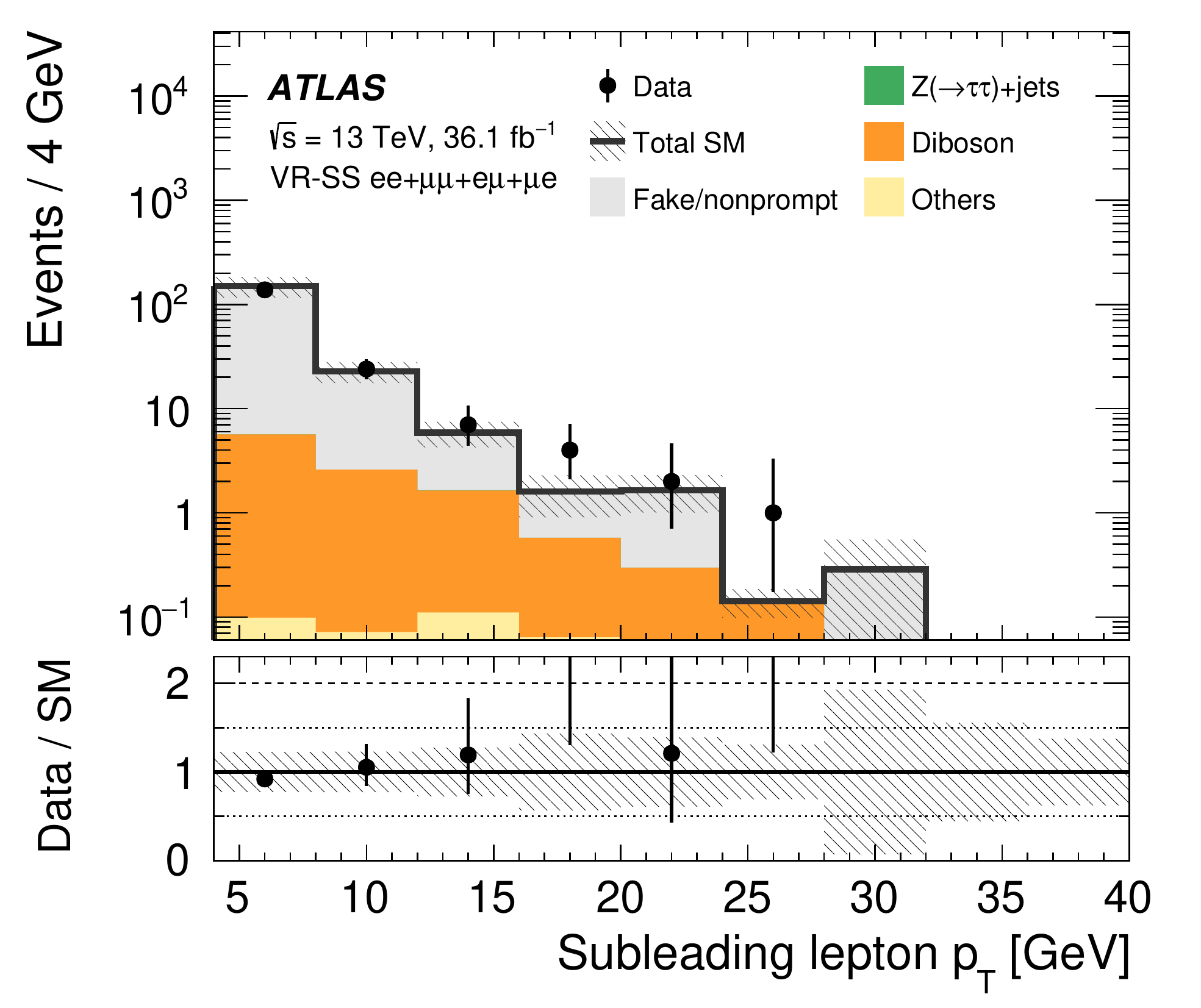}  
  \includegraphics[width=0.49\textwidth]{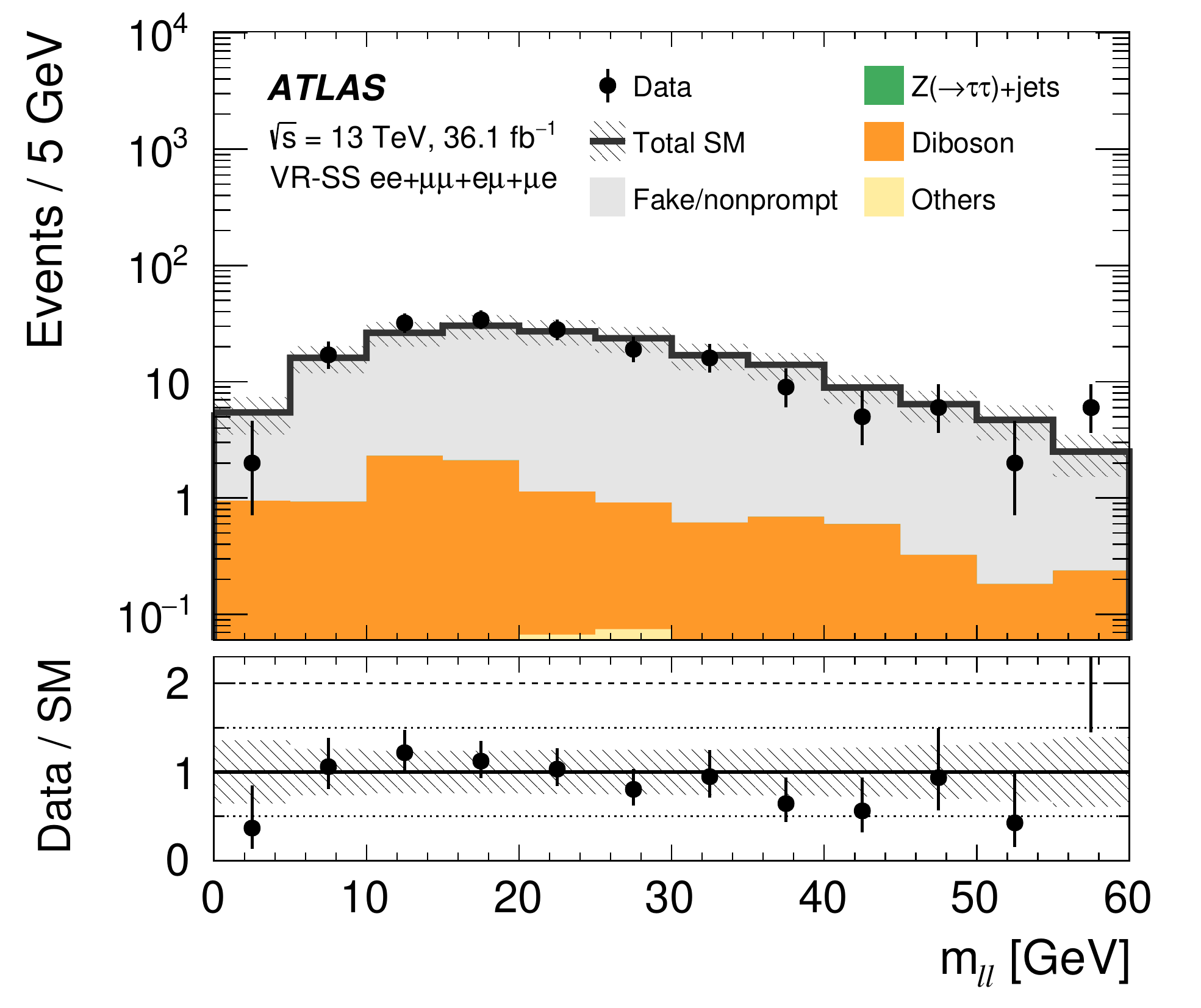}
  \end{center}
\caption{Kinematic distributions after the background-only fit showing the data and the expected background in
the different-flavor validation region VRDF-$m_\text{T2}^{100}$ (top left), 
the diboson validation region VR-VV (top right),
and the same-sign validation region VR-SS inclusive of lepton flavor (bottom).
Similar levels of agreement are observed in other kinematic distributions for VR-SS and VR-VV.
Background processes containing fewer than two prompt leptons are categorized as `Fake/nonprompt'.
The category
`Others' contains rare backgrounds from triboson, Higgs boson, and the remaining top-quark 
production processes listed in Table~\ref{tab:MCconfig}.  The last bin includes
overflow. The uncertainty bands plotted include all statistical and systematic uncertainties. Orange
arrows in the Data/SM panel indicate values that are beyond the y-axis range.}
 \label{sec:results:postfitplots}
\end{figure}

%--------------------------------------------------------------------------
The accuracy of the background
predictions is tested in the VRs discussed in
Section~\ref{sec:background}. As illustrated in
Figure~\ref{fig:pull_plot_summary_yields:vrs}, the background predictions in
the VRs are in good agreement with the observed data yields (deviations < $1.5
\sigma$).
Figure~\ref{sec:results:postfitplots} shows
distributions of the data and the expected backgrounds for a selection of
VRs and kinematic variables including the $\mll$ distribution in VR-VV 
and the $\mtt^{100}$ distribution in VR-SS. Data and background predictions are
compatible within uncertainties. 

%--------------------------------------------------------------------------

\begin{figure}[tp]
 \begin{center}
  \includegraphics[width=0.49\textwidth]{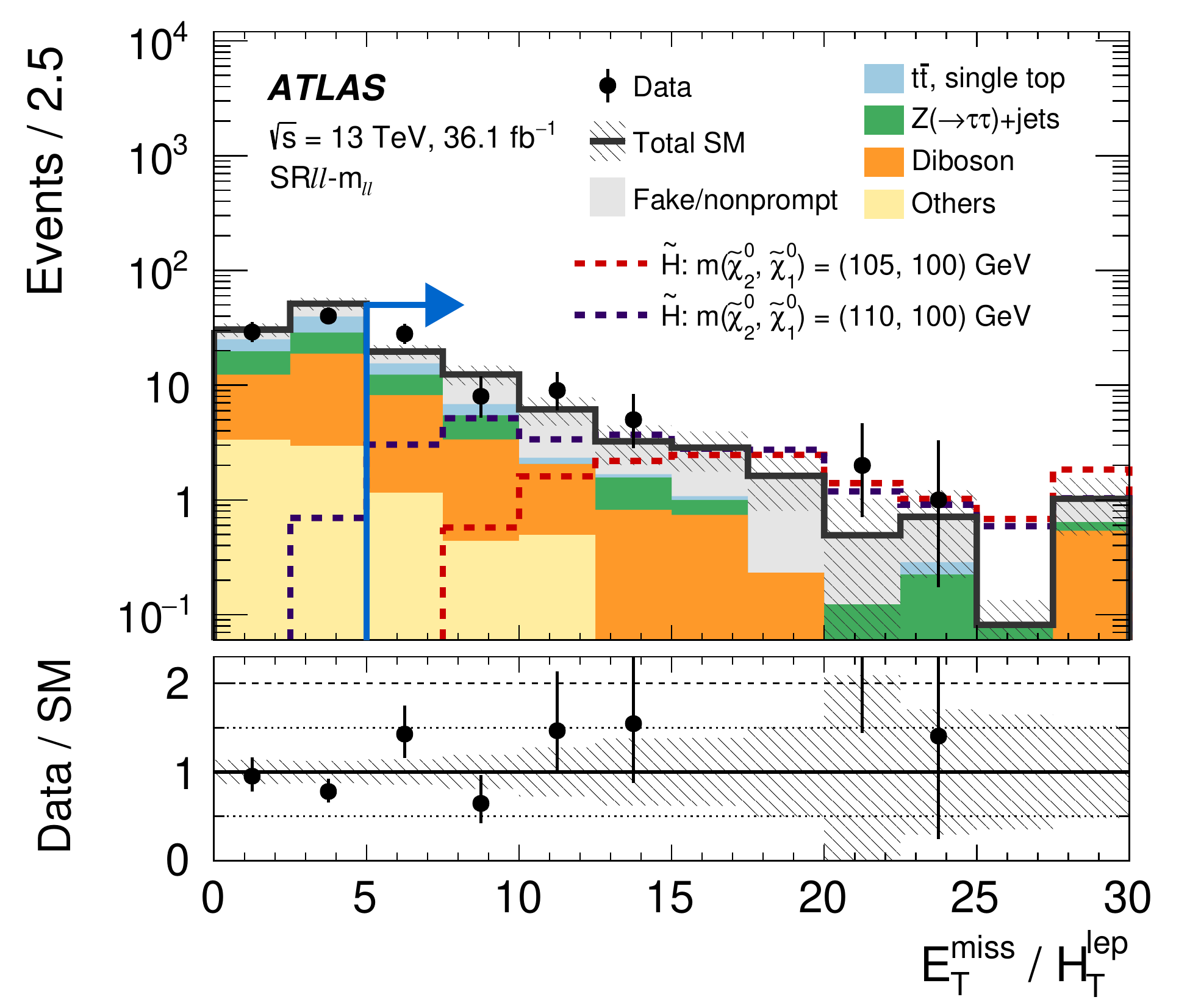}  
  \includegraphics[width=0.49\textwidth]{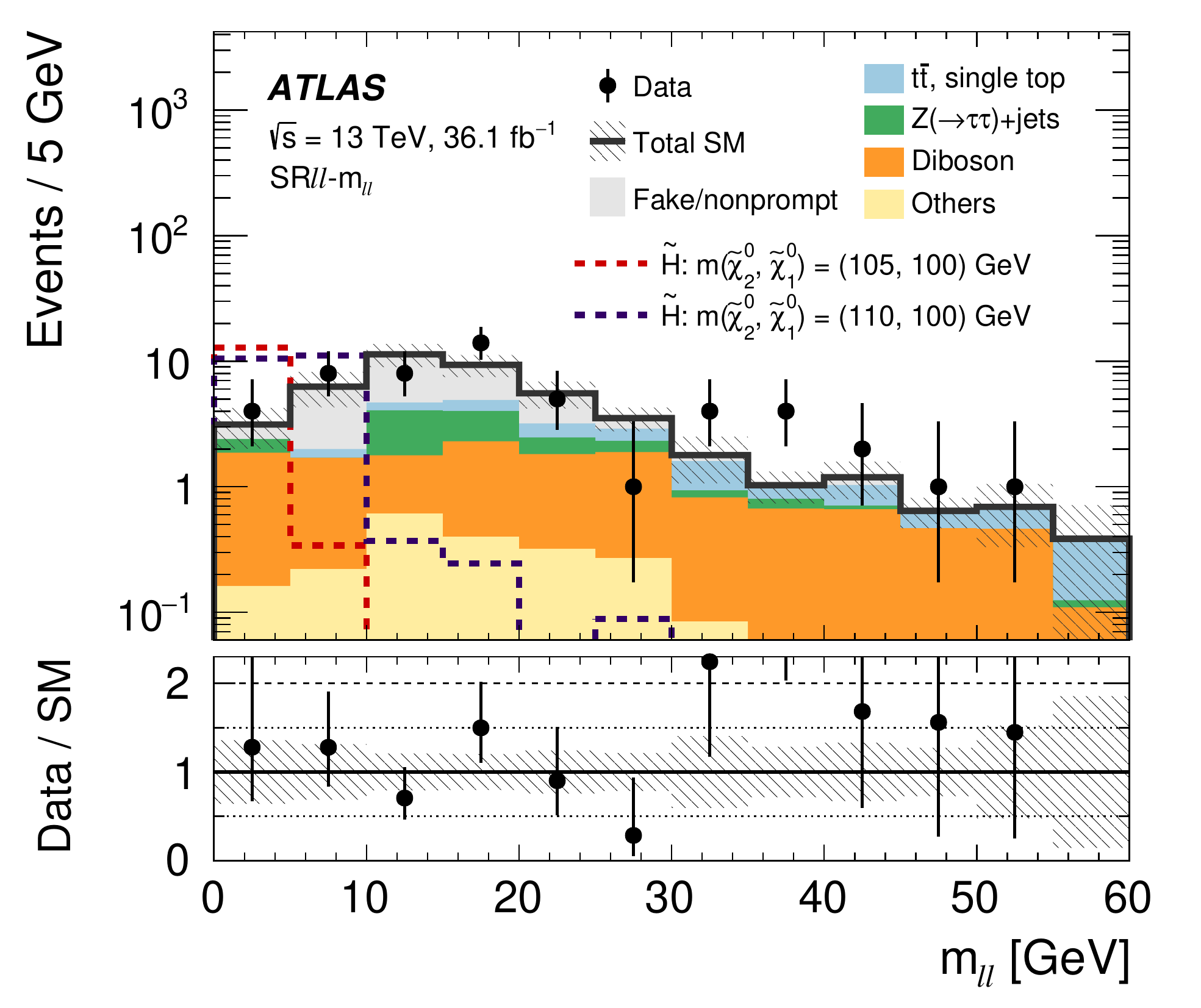}  
  \includegraphics[width=0.49\textwidth]{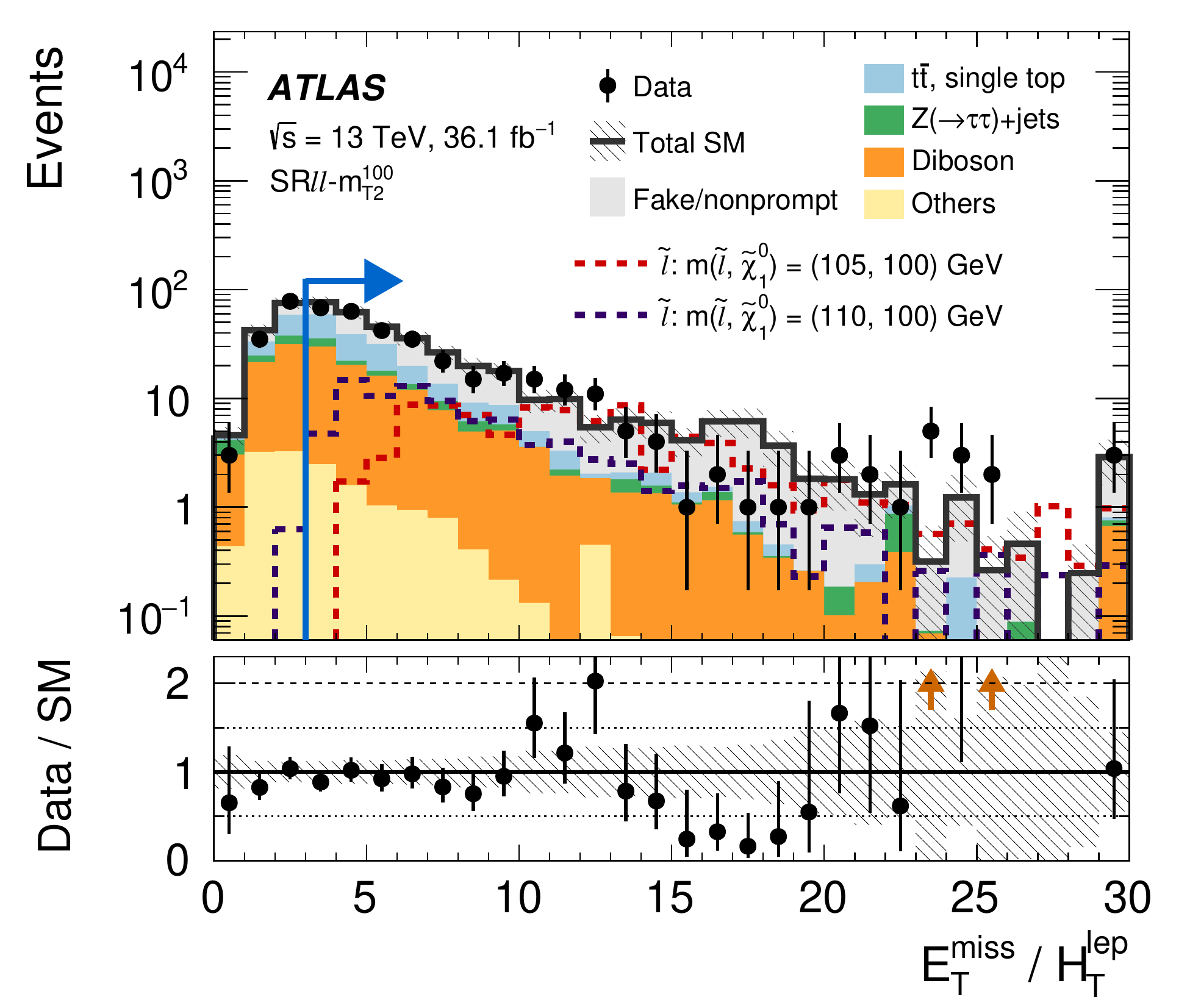}  
  \includegraphics[width=0.49\textwidth]{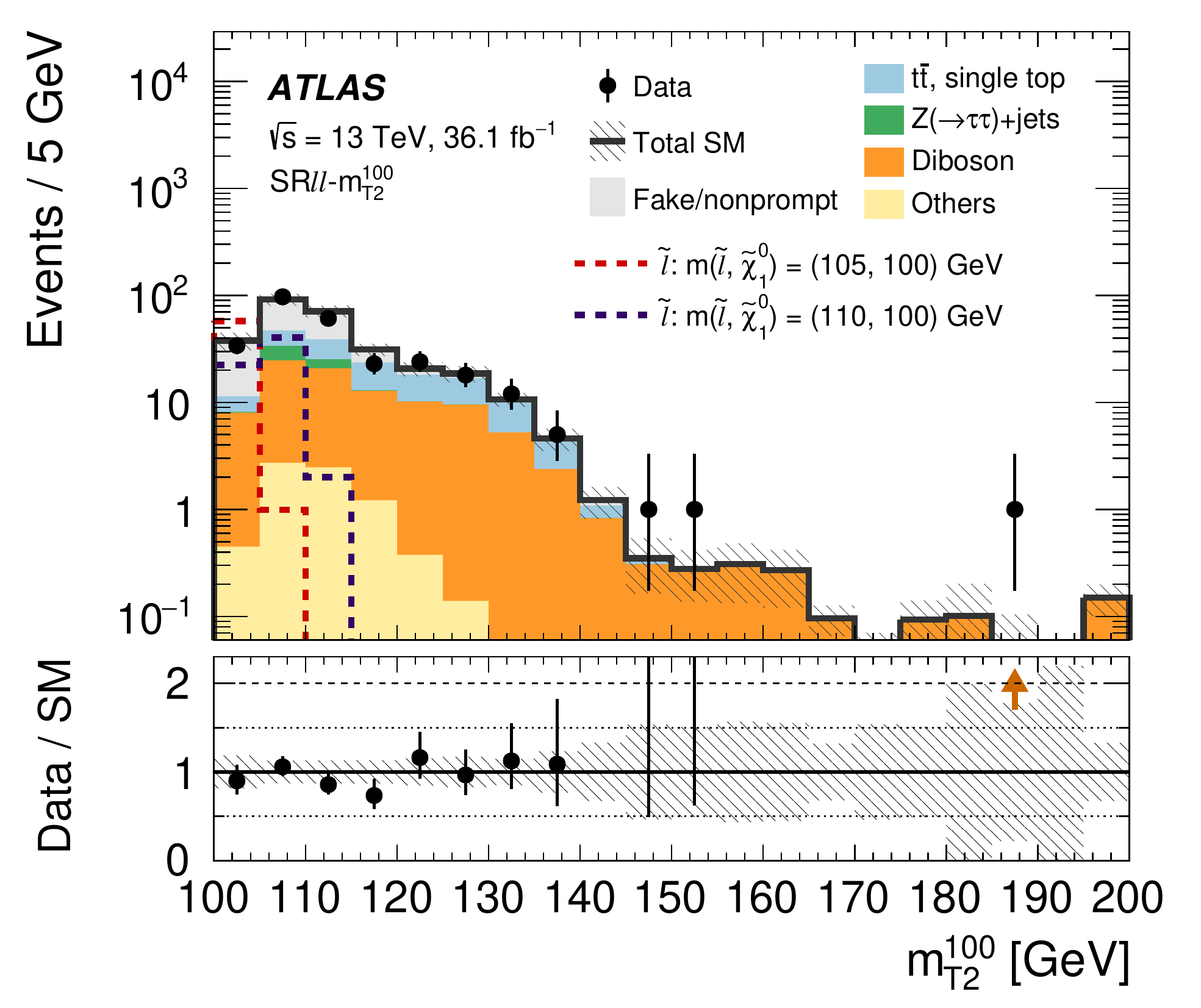}  
  \end{center}
\caption{Kinematic distributions after the background-only fit showing the data as well as the
expected background in the inclusive electroweakino \SRllmll~$[1, 60]$ (top) and
slepton \SRllmtt~$[100, \infty]$ (bottom) signal regions. 
The arrow in the $\met/\Htlep$ variables indicates the minimum value of the requirement imposed in the final SR selection. 
The $\mll$ and $\mtt$ distributions (right) have all the SR requirements applied.
Background processes containing fewer than two prompt leptons are categorized as `Fake/nonprompt'.
The category
`Others' contains rare backgrounds from triboson, Higgs boson, and the remaining top-quark
production processes listed in Table~\ref{tab:MCconfig}. 
The uncertainty bands plotted include all statistical and systematic uncertainties. 
The last bin includes overflow.
The dashed lines represent benchmark signal
samples corresponding to the Higgsino $\widetilde{H}$ and slepton $\slepton$
simplified models. 
Orange arrows in the Data/SM panel indicate values that are beyond the y-axis range.
}
 \label{sec:results:SRpostfitplots}
\end{figure}

%--------------------------------------------------------------------------
Figure~\ref{sec:results:SRpostfitplots}~shows kinematic
distributions of the data and the expected backgrounds for the inclusive SRs. 
No significant excesses of the data above the expected background are observed.

The observed and predicted event yields from the background-only fit are used to set 
model-independent upper limits on processes beyond the SM 
by including one inclusive SR at a time in a simultaneous fit with the CRs.
Using the CL$_\text{s}$ prescription~\cite{Read:2002hq}, a hypothesis test is performed to set 
upper limits at the 95\% confidence level (CL) 
on the observed (expected) number of signal events $S_\text{obs (exp)}^{95}$
in each SR.
Dividing $S_\text{obs}^{95}$ by the integrated luminosity of 36.1~fb$^{-1}$ defines the
upper limits on the visible cross-sections $\langle\epsilon\sigma\rangle_\text{
obs}^{95}$.
These results are shown in
Table~\ref{table.results.exclxsec.pval.upperlimit.SRSF_iMLLa}. To quantify the probability under the background-only hypothesis to produce event yields greater than or equal to
the observed data the discovery p-values are given as well. 

{\renewcommand{\arraystretch}{1.2}
\begin{table}
\centering
\caption[Breakdown of yields and upper limits.]{
 Left to right: The first two columns present observed ($N_{\mathrm{obs}}$) and expected ($N_{\mathrm{exp}}$) event yields in the inclusive signal regions.
 The latter are obtained by the background-only fit of the control regions, and the errors include both statistical and systematic uncertainties.
 The next two columns show the observed 95\% CL upper limits on the visible cross-section
 $\left(\langle\epsilon\sigma\rangle_\text{obs}^{95}\right)$ and on the number of
 signal events $\left(S_\text{obs}^{95}\right)$.  The fifth column
 $\left(S_\text{exp}^{95}\right)$ shows what the 95\% CL upper limit on the number of signal events would be, given an observed number of events equal to the expected number (and $\pm1\sigma$ deviations from the expectation) of background events.
 The last column indicates the discovery $p$-value ($p(s = 0)$), which is capped at 0.5.
 \label{table.results.exclxsec.pval.upperlimit.SRSF_iMLLa}
}
\vspace{0.2cm}
\renewcommand{\arraystretch}{1.2}
\setlength{\tabcolsep}{0.0pc}

\begin{tabular*}{\textwidth}{@{\extracolsep{\fill}}lrS@{\hskip -0.6pc $\pm$\hskip -1.1pc}SSSccc}
\toprule
\textbf{Signal Region}        & $N_\text{obs}$   & \multicolumn{2}{c}{ { $N_\text{exp}$} }  & { $\langle\epsilon\mathrm{\sigma}\rangle_\text{obs}^{95}$ [fb] } &  { $S_\text{obs}^{95}$ }  &  $S_\text{exp}^{95}$ & $p(s=0)$   \\
\midrule  

SR$\ell\ell$-$m_{\ell\ell}$ $[1, 3]$  & 1 & 1.7 & 0.9 & 0.10 &  3.8 & $ { 4.3 }^{ +1.7 }_{ -0.7 }$ & $0.50$  \\%
SR$\ell\ell$-$m_{\ell\ell}$ $[1, 5]$  & 4 & 3.1 & 1.2 & 0.18 &  6.6 & $ { 5.6 }^{ +2.3 }_{ -1.0 }$ & $0.32$  \\%
SR$\ell\ell$-$m_{\ell\ell}$ $[1, 10]$  & 12 & 8.9 & 2.5 & 0.34 &  12.3 & $ { 9.6 }^{ +3.2 }_{ -1.9 }$ & $0.21$  \\%
SR$\ell\ell$-$m_{\ell\ell}$ $[1, 20]$  & 34 & 29 & 6 & 0.61 &  22 & $ { 17 }^{ +7 }_{ -6 }$ & $0.25$  \\%
SR$\ell\ell$-$m_{\ell\ell}$ $[1, 30]$  & 40 & 38 & 6 & 0.59 &  21 & $ { 20 }^{ +9 }_{ -5 }$ & $0.38$  \\%
SR$\ell\ell$-$m_{\ell\ell}$ $[1, 40]$  & 48 & 41 & 7 & 0.72 &  26 & $ { 20 }^{ +8 }_{ -5 }$ & $0.20$  \\%
SR$\ell\ell$-$m_{\ell\ell}$ $[1, 60]$  & 52 & 43 & 7 & 0.80 &  29 & $ { 24 }^{ +5 }_{ -10 }$ & $0.18$  \\%
\midrule
SR$\ell\ell$-$m_\text{T2}^{100}$ $[100, 102]$  & 8 & 12.4 & 3.1 & 0.18 &  7 & $ { 9 }^{ +4 }_{ -2 }$ & $0.50$  \\%
SR$\ell\ell$-$m_\text{T2}^{100}$ $[100, 105]$  & 34 & 38 & 7 & 0.49 &  18 & $ { 23 }^{ +7 }_{ -7 }$ & $0.50$  \\%
SR$\ell\ell$-$m_\text{T2}^{100}$ $[100, 110]$  & 131 & 129 & 18 & 1.3 &  48 & $ { 47 }^{ +13 }_{ -15 }$ & $0.37$  \\%
SR$\ell\ell$-$m_\text{T2}^{100}$ $[100, 120]$  & 215 & 232 & 29 & 1.4 &  52 & $ { 62 }^{ +21 }_{ -15 }$ & $0.50$  \\%
SR$\ell\ell$-$m_\text{T2}^{100}$ $[100, 130]$  & 257 & 271 & 32 & 1.7 &  61 & $ { 69 }^{ +22 }_{ -17 }$ & $0.50$  \\%
SR$\ell\ell$-$m_\text{T2}^{100}$ $[100, \infty]$  & 277 & 289 & 33 & 1.8 &  66 & $ { 72 }^{ +24 }_{ -17 }$ & $0.50$  \\%

\bottomrule
\end{tabular*}
\end{table}
}
%--------------------------------------------------------------------------

\FloatBarrier

In the absence of any significant deviations from the SM expectation
in the inclusive SRs, the results are interpreted as constraints on
the SUSY models discussed in Section~\ref{sec:samples} using the
exclusive electroweakino and slepton SRs. The background-only fit is extended
to allow for a signal model with a corresponding signal strength parameter in a
simultaneous fit of all CRs and relevant SRs; this is referred to
as the exclusion fit.
When an electroweakino signal is assumed, the 14 exclusive \SReemll\ and
\SRmmmll\ regions binned in \mll\ are considered.
By statistically combining these SRs, the signal shape of the
\mll\ spectrum can be exploited to improve the sensitivity.
When a slepton signal is assumed, the 12 exclusive \SReemtt\ and \SRmmmtt\
regions binned in \mtth\ are used for the fit.

\begin{figure}
    \centering
        \includegraphics[width=\textwidth]{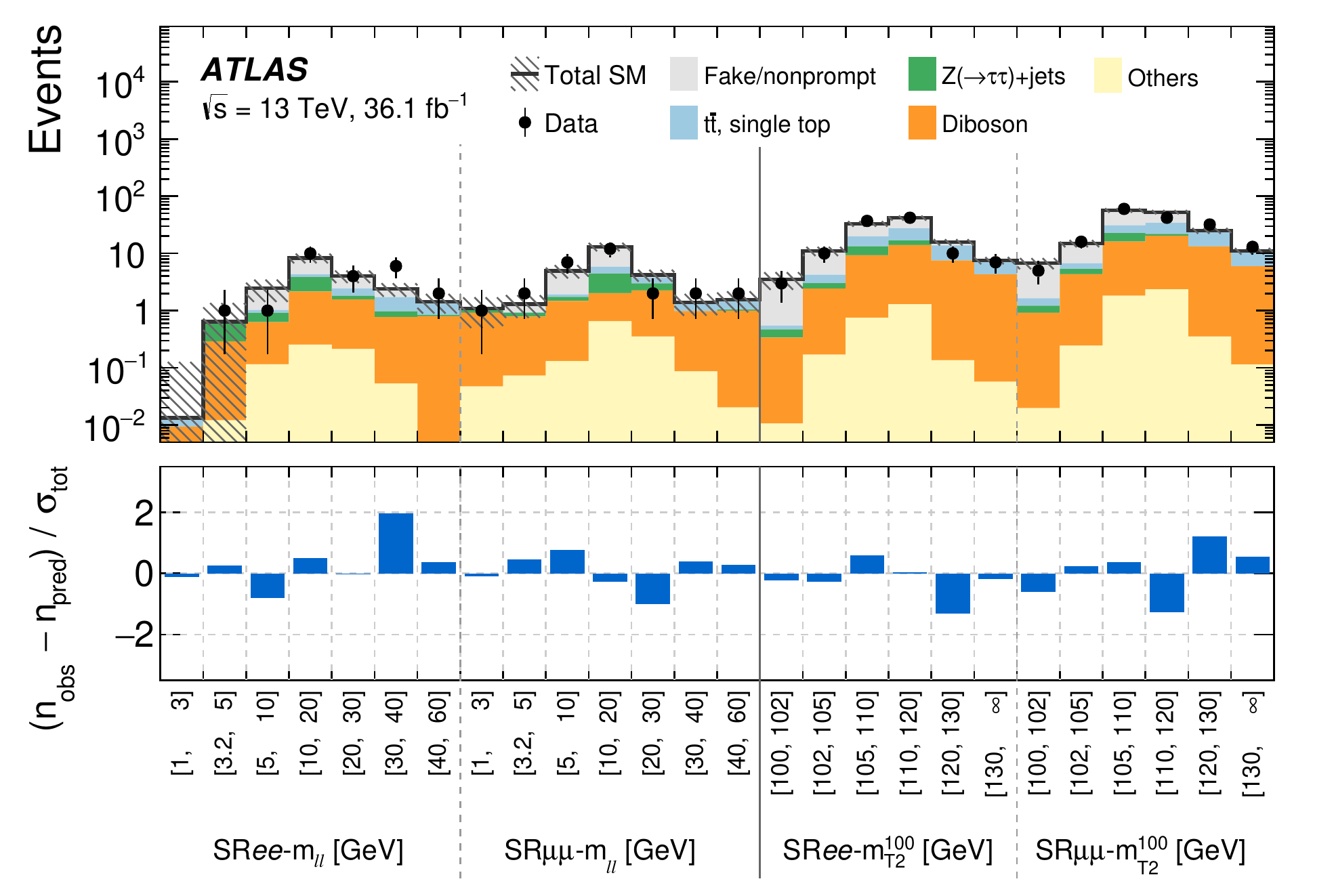}
        \caption{Comparison of observed and expected event yields after the
        exclusion fit with the signal strength parameter set to zero in the exclusive signal regions. 
       Background processes containing fewer than two prompt leptons are categorized as `Fake/nonprompt'.
        The category
`Others' contains rare backgrounds from triboson, Higgs boson, and the remaining top-quark
 production processes listed in Table~\ref{tab:MCconfig}. Uncertainties in the background estimates include
    both the statistical and systematic uncertainties, where $\sigma_\text{tot}$ denotes the total uncertainty.}
    \label{fig:pull_plot_summary_yields:exclSRs}
\end{figure}

Table~\ref{tab:results:exclusiveSRYields} summarizes the fitted and observed
event yields in the exclusive electroweakino and slepton SRs using an
exclusion fit configuration where the signal strength parameter is fixed to
zero.
The predicted yields differ slightly from those obtained in the background-only fit,
as expected, because inclusion of the SRs to the fit further constrains the
background contributions in the absence of signal.
Figure~\ref{fig:pull_plot_summary_yields:exclSRs} illustrates the
compatibility of the fitted and observed event yields in these regions. No
significant differences between the fitted background and 
the observed event yields are found in
the exclusive SRs.

\begin{table}
\caption{Observed event yields and exclusion fit results with the signal
strength parameter set to zero for the exclusive electroweakino and slepton signal
regions. 
Background processes containing fewer than two prompt leptons are categorized as `Fake/nonprompt'.
The category
`Others' contains rare backgrounds from triboson, Higgs boson, and the remaining top-quark
production processes listed in Table~\ref{tab:MCconfig}. Uncertainties in the fitted background estimates combine statistical and
systematic uncertainties.}
\label{tab:results:exclusiveSRYields}
\begin{center}
\setlength{\tabcolsep}{0.0pc}
{\scriptsize
\renewcommand{\arraystretch}{1.2}
\begin{tabular*}{\textwidth}{@{\extracolsep{\fill}}lccccccc}
\toprule
SR$ee$-$m_{\ell\ell}$            & [1, 3] \GeV            & [3.2, 5] \GeV            & [5, 10] \GeV            & [10, 20] \GeV            & [20, 30] \GeV            & [30, 40] \GeV            & [40, 60] \GeV              \\[-0.05cm]
\midrule
Observed events          & $0$              & $1$              & $1$              & $10$              & $4$              & $6$              & $2$                    \\
\midrule\midrule
Fitted SM events         & $0.01_{-0.01}^{+0.11}$ & $0.6_{-0.6}^{+0.7}$ & $2.4 \pm 1.0$     & $8.3 \pm 1.6$     & $4.0 \pm 1.0$     & $2.4 \pm 0.6$     & $1.4 \pm 0.       5$         \\
\midrule
Fake/nonprompt leptons & $0.00_{-0.00}^{+0.08}$   & $0.02_{-0.02}^{+0.12}$ & $1.4 \pm 0.9$     & $4.0 \pm 1.5$     & $1.6 \pm 0.9$     & $0.7 \pm 0.6$     & $0.02_{-0.02}^{+ 0.11}$     \\
Diboson & $0.007_{-0.007}^{+0.014}$ & $0.28_{-0.28}^{+0.29}$ & $0.51 \pm 0.28$     & $1.9 \pm 0.6$     & $1.36 \pm 0.31$     & $0.72 \pm 0.22$     & $0.80 \pm 0.           28$         \\
$Z(\to\tau\tau)$+jets & $0.000_{-0.000}^{+0.007}$ & $0.3_{-0.3}^{+0.8}$ & $0.3_{-0.3}^{+0.5}$ & $1.7 \pm 0.7$     & $0.25_{-0.25}^{+0.26}$ & $0.20 \pm 0.18$     & $0.04_{- 0.04}^{+0.28}$     \\
$t\bar{t}$, single top & $0.00_{-0.00}^{+0.08}$ & $0.02_{-0.02}^{+0.12}$ & $0.11_{-0.11}^{+0.14}$ & $0.44 \pm 0.29$     & $0.63 \pm 0.35$     & $0.7 \pm 0.4$     & $0.6    \pm 0.4$         \\
Others & $0.002_{-0.002}^{+0.015}$ & $0.012_{-0.012}^{+0.013}$ & $0.12 \pm 0.11$     & $0.25 \pm 0.16$     & $0.21 \pm 0.12$     & $0.05_{-0.05}^{+0.06}$ & $0.0018_{-0.    0018}^{+0.0033}$     \\
 
\midrule\midrule

SR$\mu\mu$-$m_{\ell\ell}$            & [1, 3] \GeV            & [3.2, 5] \GeV            & [5, 10] \GeV            & [10, 20] \GeV            & [20, 30] \GeV            & [30, 40] \GeV            & [40, 60] \GeV              \\[-0.05cm]
\midrule
Observed events          & $1$              & $2$              & $7$              & $12$              & $2$              & $2$              & $2$                    \\
\midrule\midrule
Fitted SM events         & $1.1 \pm 0.6$     & $1.3 \pm 0.6$     & $4.9 \pm 1.3$     & $13.1 \pm 2.2$     & $4.2 \pm 1.0$     & $1.4 \pm 0.6$     & $1.6 \pm 0.6$         \\
\midrule
Fake/nonprompt leptons & $0.00_{-0.00}^{+0.33}$     & $0.4_{-0.4}^{+0.5}$ & $3.0 \pm 1.3$     & $7.3 \pm 2.1$     & $0.4_{-0.4}^{+0.8}$ & $0.03_{-0.03}^{+0.19}$ & $0.0_{-0.0}^{+0.5}$         \\
Diboson & $0.9 \pm 0.5$     & $0.7 \pm 0.4$     & $1.3 \pm 0.6$     & $1.4 \pm 0.5$     & $1.9 \pm 0.4$     & $0.9 \pm 0.5$     & $0.97 \pm 0.28$         \\
$Z(\to\tau\tau)$+jets & $0.18_{-0.18}^{+0.25}$ & $0.13 \pm 0.12$     & $0.3_{-0.3}^{+0.5}$ & $2.4 \pm 0.8$     & $0.7 \pm 0.4$     & $0.001_{-0.001}^{+0.011}$ & $0.05_{-0. 05}^{+0.06}$     \\
$t\bar{t}$, single top & $0.01_{-0.01}^{+0.10}$ & $0.02_{-0.02}^{+0.12}$ & $0.19 \pm 0.13$     & $1.4 \pm 0.6$     & $0.8 \pm 0.4$     & $0.37 \pm 0.21$     & $0.51 \pm 0. 33$         \\
Others & $0.047 \pm 0.030$     & $0.07_{-0.07}^{+0.09}$ & $0.13 \pm 0.12$     & $0.7 \pm 0.5$     & $0.35 \pm 0.20$     & $0.09 \pm 0.07$     & $0.020 \pm 0.020$     \\

\midrule

\vspace*{0.5cm}\\

\midrule

SR$ee$-$m_\text{T2}^{100}$          & & [100, 102] \GeV            & [102, 105] \GeV            & [105, 110] \GeV            & [110, 120] \GeV            & [120, 130] \GeV            & $[130, \infty]$ \GeV              \\[-0.05cm]
\midrule
Observed events         & & $3$              & $10$              & $37$              & $42$              & $10$              & $7$                    \\
\midrule\midrule
Fitted SM events        & & $3.5 \pm 1.2$     & $11.0 \pm 2.0$     & $33 \pm 4$     & $42 \pm 4$     & $15.7 \pm 2.0$     & $7.5 \pm 1.1$         \\
\midrule
Fake/nonprompt leptons & & $2.9 \pm 1.2$     & $6.8 \pm 2.0$     & $13 \pm 4$     & $14 \pm 4$     & $1.9 \pm 1.2$     & $0.01_{-0.01}^{+0.10}$     \\
Diboson & & $0.33 \pm 0.12$     & $2.3 \pm 0.6$     & $8.5 \pm 1.6$     & $12.7 \pm 2.4$     & $7.4 \pm 1.4$     & $4.3 \pm 0.9$         \\
$Z(\to\tau\tau)$+jets & & $0.13_{-0.13}^{+0.23}$ & $0.6 \pm 0.4$     & $4.1 \pm 1.8$     & $2.9 \pm 1.0$     & $0.00_{-0.00}^{+0.08}$ & $0.00_{-0.00}^{+0.20}$     \\
$t\bar{t}$, single top & & $0.08 \pm 0.08$ & $1.2 \pm 0.5$     & $6.5 \pm 1.6$     & $10.7 \pm 2.4$     & $6.3 \pm 1.4$     & $3.2 \pm 0.9$         \\
Others & & $0.011_{-0.011}^{+0.012}$ & $0.17 \pm 0.11$     & $0.8 \pm 0.4$     & $1.3 \pm 0.7$     & $0.14 \pm 0.09$     & $0.06 \pm 0.04$         \\

\midrule\midrule

SR$\mu\mu$-$m_\text{T2}^{100}$           & & [100, 102] \GeV            & [102, 105] \GeV            & [105, 110] \GeV            & [110, 120] \GeV            & [120, 130] \GeV            & $[130, \infty]$ \GeV              \\[-0.05cm]
\midrule
Observed events          & & $5$              & $16$              & $60$              & $42$              & $32$              & $13$                    \\
\midrule\midrule
Fitted SM events         & & $6.8 \pm 1.5$     & $15.0 \pm 2.1$     & $57 \pm 5$     & $53 \pm 4$     & $24.9 \pm 2.9$     & $11.0 \pm 1.4$         \\
\midrule
Fake/nonprompt leptons & & $5.1 \pm 1.5$     & $8.2 \pm 2.1$     & $26 \pm 5$     & $18 \pm 4$     & $1.2 \pm 0.8$     & $0.02_{-0.02}^{+0.17}$     \\
Diboson & & $0.89 \pm 0.22$     & $4.1 \pm 0.9$     & $14.3 \pm 2.2$     & $18.0 \pm 2.7$     & $12.9 \pm 2.2$     & $5.9 \pm 1.1$         \\
$Z(\to\tau\tau)$+jets & & $0.31 \pm 0.23$     & $1.0_{-1.0}^{+1.3}$ & $6.6 \pm 1.7$     & $1.6_{-1.6}^{+1.8}$ & $0.03_{-0.03}^{+0.25}$ & $0.02_{-0.02}^{+0.24}$     \\
$t\bar{t}$, single top & & $0.43 \pm 0.22$     & $1.4 \pm 0.5$     & $8.3 \pm 2.2$     & $12.4 \pm 2.9$     & $10.5 \pm 2.6$     & $5.0 \pm 1.3$         \\
Others & & $0.020_{-0.020}^{+0.024}$ & $0.24 \pm 0.15$     & $1.8 \pm 1.0$     & $2.4 \pm 1.3$     & $0.35 \pm 0.23$     & $0.11 \pm 0.07$         \\
 
\bottomrule
\end{tabular*}
%%%
}
\end{center}
\end{table}

\begin{figure}
    \centering
    \begin{subfigure}[b]{0.85\textwidth}
      \centering
        \includegraphics[width=\columnwidth]{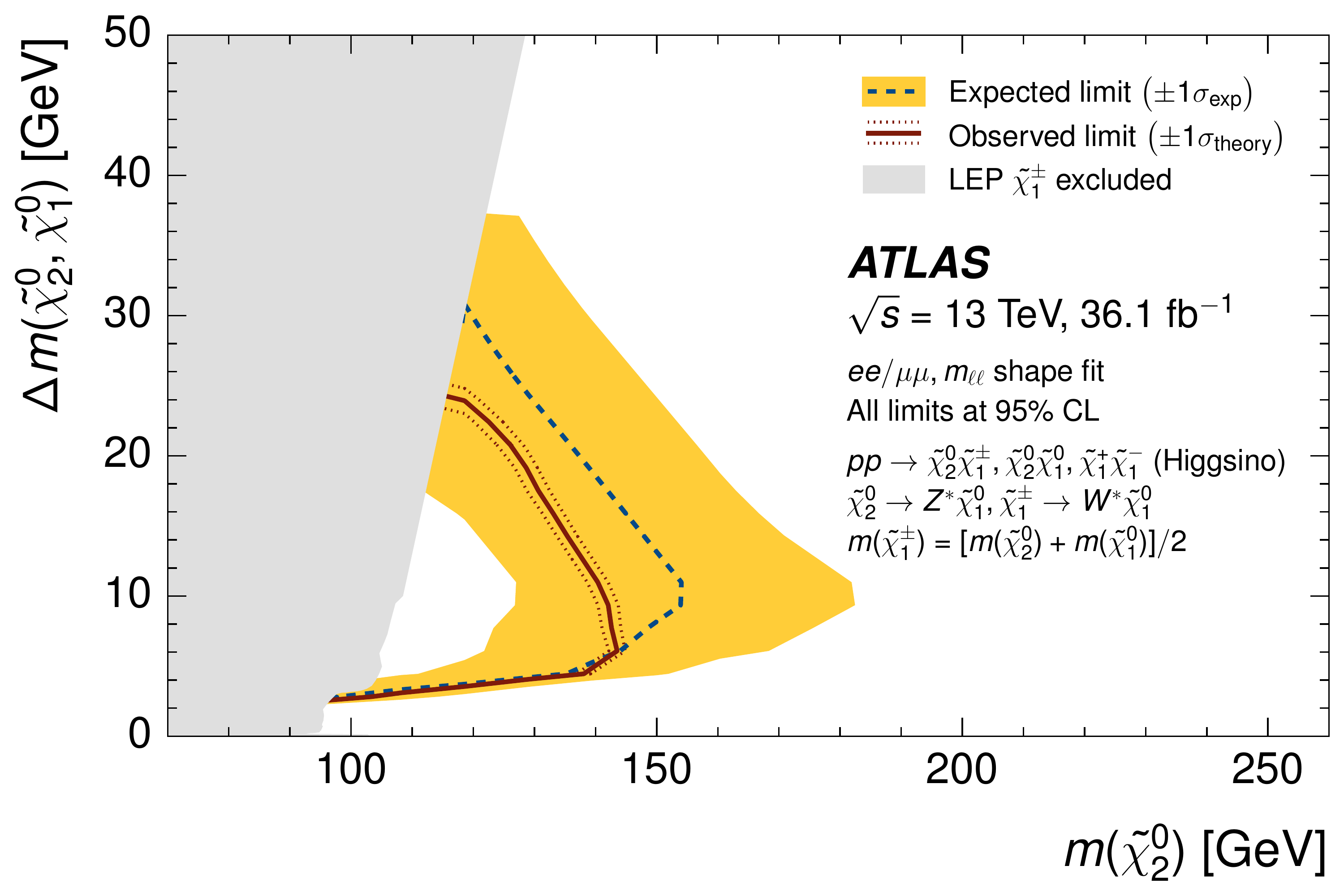}   
    \end{subfigure} 
    \begin{subfigure}[b]{0.85\textwidth}
      \centering
        \includegraphics[width=\columnwidth]{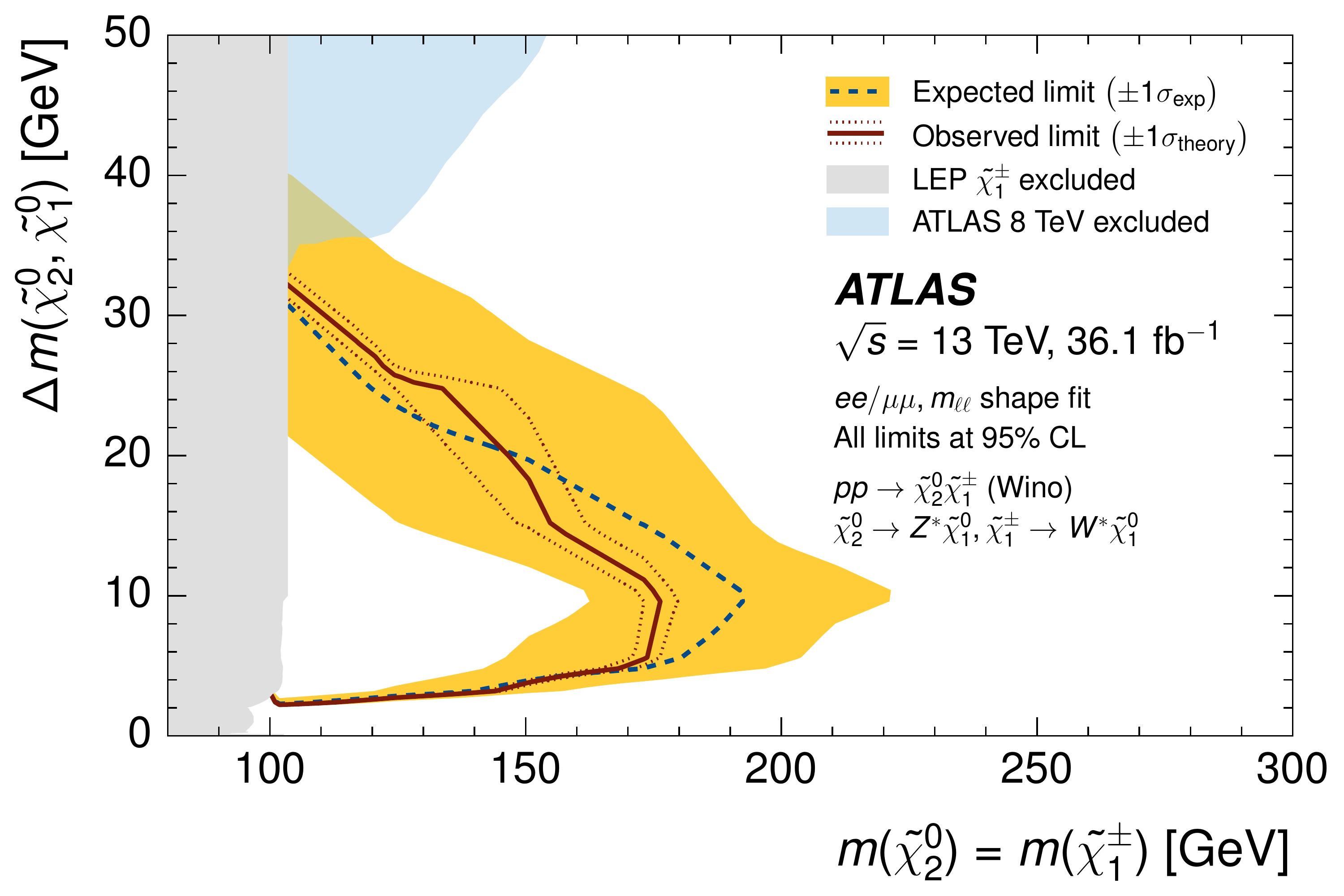}   
    \end{subfigure} 
    \caption{
    Expected 95\% CL exclusion sensitivity (blue dashed line) with
    $\pm1\sigma_\text{exp}$ (yellow band) from experimental systematic uncertainties and
    observed limits (red solid line) with $\pm1\sigma_\text{theory}$ (dotted red line)
    from signal cross-section uncertainties for simplified models of
    direct Higgsino (top) and wino (bottom) production. A fit of signals
    to the \mll\ spectrum is used to derive the limit, 
    which is projected into the $\Delta m(\chitz, \chioz)$ vs.\ $m(\chitz)$ plane. For Higgsino production, the chargino
    $\chiopm$ mass is assumed to be halfway between the two lightest
    neutralino masses, while $m(\chitz) = m(\chiopm)$ is assumed for the wino--bino
    model.
    The gray regions denote the lower chargino mass limit from
    LEP~\cite{LEPlimits}.
    The blue region in the lower plot indicates the limit from the $2\ell+3\ell$
    combination of ATLAS Run~1~\mbox{\cite{SUSY-2013-11,SUSY-2013-12}.}}
\label{fig:exclusion_contour_higgsinowino}
\end{figure}

Hypothesis tests are then performed to set limits 
on simplified model scenarios using the CL$_\text{s}$ prescription.
Figure~\ref{fig:exclusion_contour_higgsinowino} (top) shows the 95\% CL limits
on the Higgsino simplified model, based on an exclusion fit that
exploits the shape of the \mll\ spectrum using the exclusive electroweakino
SRs.
The exclusion limits are projected into the next-to-lightest neutralino mass
$\Delta m(\chitz,\chioz)$ versus $m(\chitz)$ plane, where $\chitz$ are excluded up
to masses of $\sim$145~\GeV{} for $\Delta m(\chitz,\chioz)$ between 5~\GeV{} and
10~\GeV, and down to $\Delta m(\chitz,\chioz) \sim 2.5~\GeV$ for $m(\chitz)\sim 100~\GeV$.
The 95\% CL limits of the wino--bino simplified model are
shown in Figure~\ref{fig:exclusion_contour_higgsinowino} (bottom), 
where $\chitz$ neutralino is excluded up to masses of $\sim$175~\GeV{} 
for $\Delta m(\chitz,\chioz) \sim 10~\GeV$, and down $\Delta m(\chitz,\chioz)\sim 2~\GeV$ for $m(\chitz)\sim 100~\GeV$.

\begin{figure}
    \centering
        \includegraphics[width=0.85\textwidth]{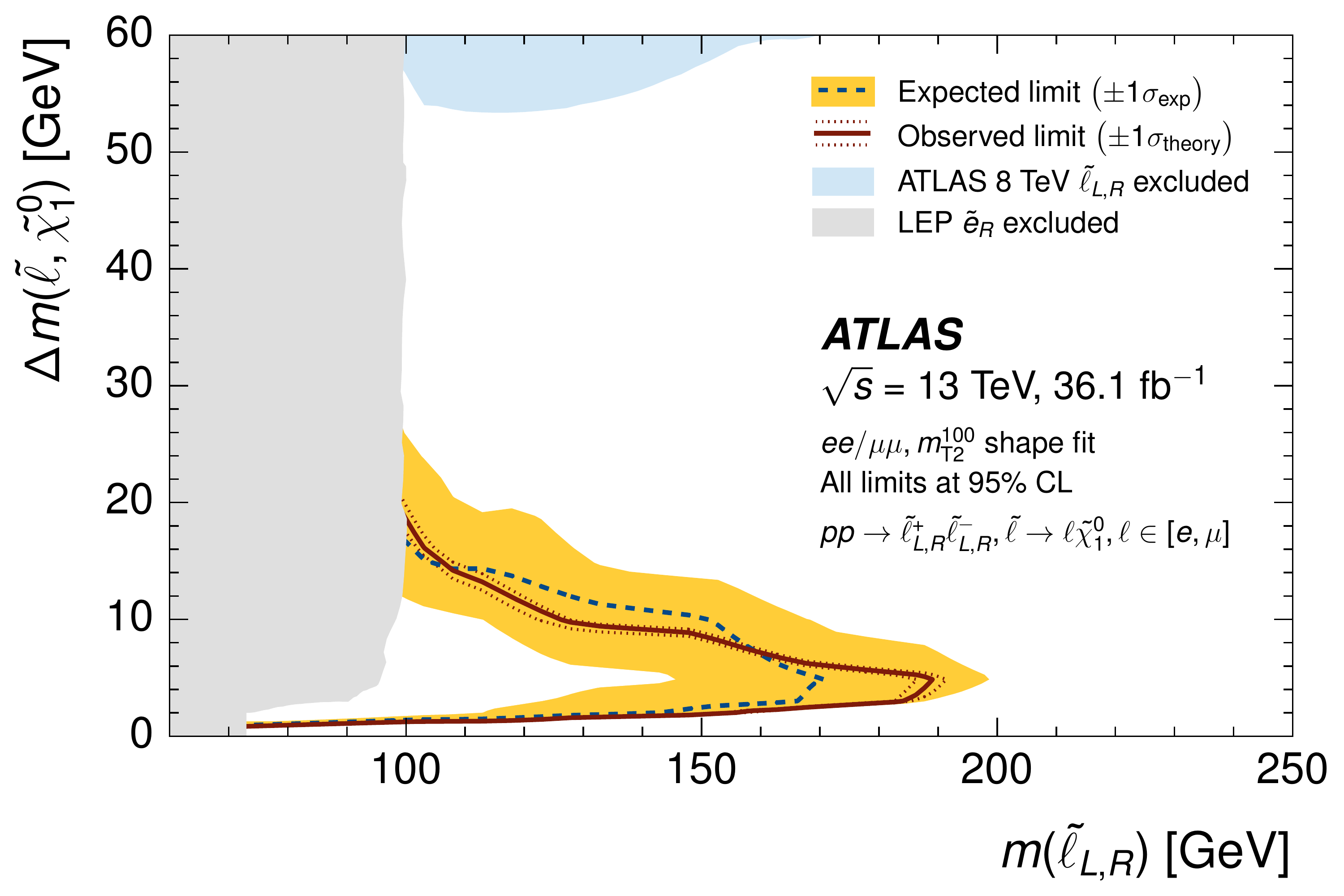}   
    \caption{
    Expected 95\% CL exclusion sensitivity (blue dashed line) with $\pm 1 \sigma_\text{exp}$ (yellow band) from experimental systematic uncertainties
    and observed limits (red solid line) with $\pm 1 \sigma_\text{theory}$ (dotted red line) from signal cross-section uncertainties
    for simplified models of direct slepton production. 
    A fit of slepton signals to the \mtth\ spectrum is used to derive 
    the limit, which is projected into the $\Delta m(\slepton, \chioz)$ vs.\ $m(\slepton)$ plane. 
    Slepton $\slepton$ refers to the
    scalar partners of left- and right-handed electrons and muons,
    which are assumed to be fourfold mass degenerate
     $m(\selectron_L) = m(\selectron_R) = m(\smuon_L) = m(\smuon_R)$.
    The gray region is the $\widetilde{e}_R$ limit from LEP~\mbox{\cite{LEPlimits,Heister:2002jca}}, 
    while the blue region is the fourfold mass degenerate slepton limit from ATLAS Run 1~\cite{SUSY-2013-11}.}
    \label{fig:exclusion_contour_slepton}
\end{figure}

Figure~\ref{fig:exclusion_contour_slepton} shows the 95\% CL
limits on the slepton simplified model, based on an exclusion fit that
exploits the shape of the \mtth\ spectrum using the exclusive slepton SRs.
Here, $\slepton$ with masses of up to $\sim$190~\GeV{} are excluded
for $\Delta m(\slepton, \chioz)\sim 5~\GeV$, and down to mass splittings $\Delta m(\slepton, \chioz)$ of
approximately 1~\GeV{} for $m(\slepton)\sim 70~\GeV$.
A fourfold degeneracy is assumed in selectron and smuon masses.

\begin{figure}
    \centering
        \includegraphics[width=0.75\textwidth]{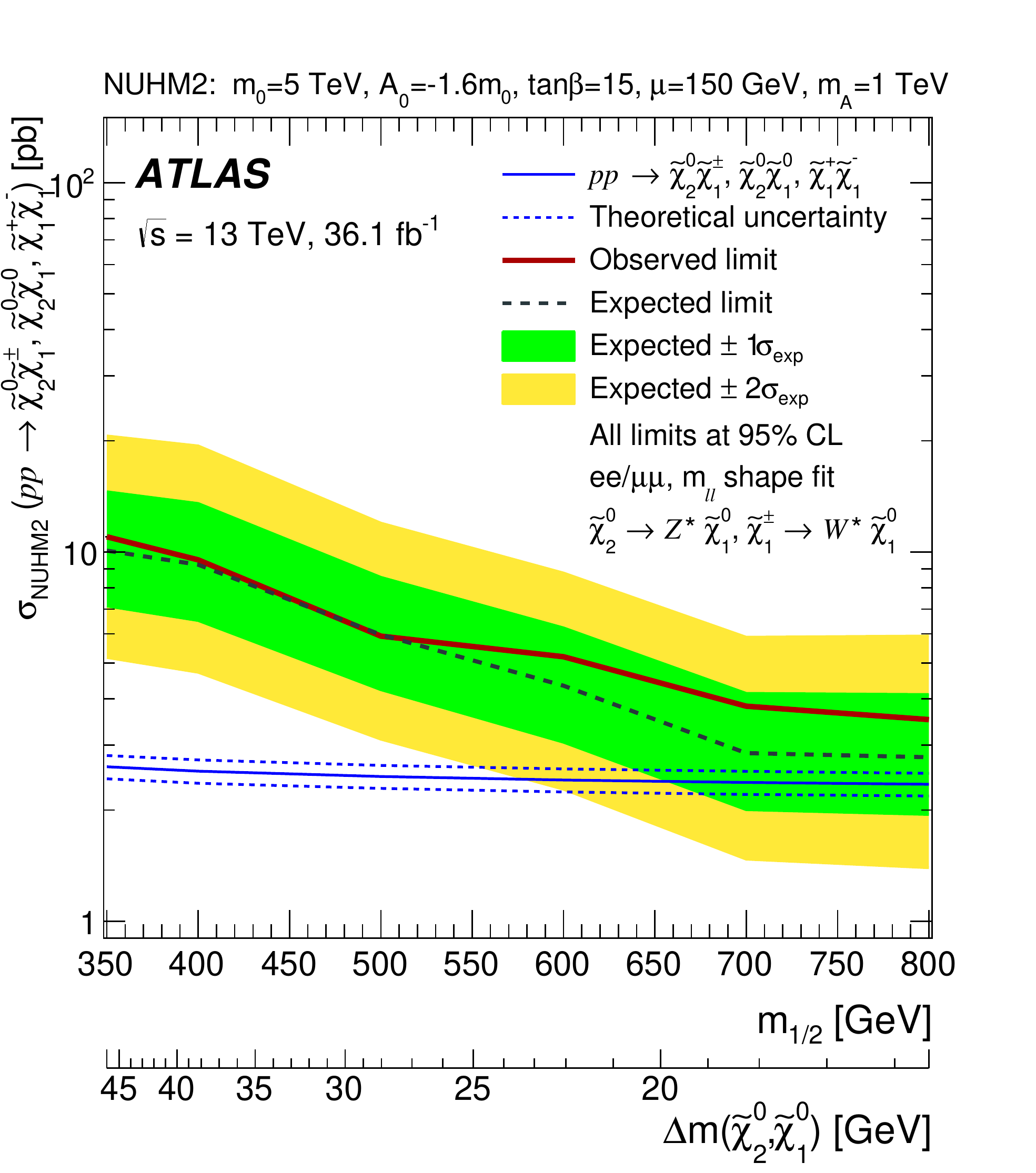}
        \caption{Observed and expected 95\% CL cross-section upper limits as a function of the universal gaugino mass $m_{1/2}$ for the NUHM2
        model.
        The green and yellow bands around the expected limit indicate the
        $\pm 1\sigma$ and $\pm 2\sigma$ uncertainties, respectively.
        The expected signal production cross-sections as well as the associated uncertainty are indicated
        with the blue solid and dashed lines.
        The lower $x$-axis indicates the difference between the \chitz{} and \chioz{} masses for different
        values of $m_{1/2}$.  A fit of signals to the \mll\ spectrum is used to derive this limit.}
    \label{fig:NUHM2}
\end{figure}

Finally, Figure~\ref{fig:NUHM2} shows the 95\% CL exclusion bounds on the
production cross-sections for the NUHM2 scenario as a function of the universal
gaugino mass $m_{1/2}$.   The NUHM2 fit exploits the shape of the \mll\ spectrum using the
exclusive electroweakino SRs.
At $m_{1/2}=350~\GeV{}$, which corresponds to a mass splitting $\Delta m(\chitz, \chioz)$ 
of approximately
45~\GeV, the signal cross-section is constrained to be less than five times the
predicted value in the NUHM2 scenario at 95\% CL. For
$m_{1/2}=800$~\GeV, corresponding to a mass splitting of approximately 15~\GeV,
the 95\% CL cross-section upper limit is twice the NUHM2
prediction.

In these interpretations, 
sensitivity is lost when the mass splitting between the produced SUSY particle and the LSP 
becomes less than a few~\GeV{} due to the reduced acceptance and reconstruction efficiency of the soft leptons. 
Meanwhile, sensitivity decreases for larger mass splittings above approximately 20 to 30~\GeV{}  
due to the \mll\ or \mtth\ shapes of the signal
becoming increasingly similar to those of the SM backgrounds.

%-------------------------------------------------------------------------------

\FloatBarrier

%-------------------------------------------------------------------------------

\section{Conclusion}
\label{sec:conclusion}
A search for the electroweak production of supersymmetric states 
with low-momentum visible decay products is performed 
using LHC proton--proton collision data collected by the ATLAS detector at $\sqrt{s}=13$~\TeV, 
corresponding to an integrated luminosity of 36.1~\ifb. 
Events with significant missing transverse momentum 
and same-flavor opposite-charge lepton pairs are selected, 
with the minimum \pt of the electrons (muons) being 4.5~(4)~\GeV.
The dilepton invariant mass and stranverse mass are the main discriminating variables used to construct signal regions. 
No excess over the Standard Model expectation is observed. 

The results are interpreted using simplified models of $R$-parity-conserving
supersymmetry, where the produced states have small mass splittings
with the lightest neutralino \chioz.
For the Higgsino simplified model,
exclusion limits at 95\% CL are set on the $\chitz$ neutralino
up to masses of $\sim$145~\GeV{} and down to mass splittings $\Delta m(\chitz,\chioz)\sim 2.5~\GeV$. 
In the wino--bino model, these limits on the
$\chitz$ extend to masses of up to $\sim$175~\GeV{} and down to mass splittings of
approximately 2~\GeV.
Direct pair production of sleptons, assuming the scalar partners 
of the left- and right-handed electrons and muons are mass degenerate,
is excluded for slepton masses up to masses of $\sim$190~\GeV{} and down to mass
splittings $\Delta m(\slepton, \chioz)\sim 1~\GeV$.
These results extend previous constraints from the LEP experiments. In addition,
an interpretation of the results in the NUHM2 scenario is provided, where the
cross-section upper limit ranges between 11 and 3.5~pb for $m_{1/2}$ values of 350 to
800~\GeV.

%-------------------------------------------------------------------------------

%-------------------------------------------------------------------------------
\section*{Acknowledgments}
% Acknowledgements for papers with collision data
% Version 29-Jan-2018

% Standard acknowledgements start here
%----------------------------------------------
We thank CERN for the very successful operation of the LHC, as well as the
support staff from our institutions without whom ATLAS could not be
operated efficiently.

We acknowledge the support of ANPCyT, Argentina; YerPhI, Armenia; ARC, Australia; BMWFW and FWF, Austria; ANAS, Azerbaijan; SSTC, Belarus; CNPq and FAPESP, Brazil; NSERC, NRC and CFI, Canada; CERN; CONICYT, Chile; CAS, MOST and NSFC, China; COLCIENCIAS, Colombia; MSMT CR, MPO CR and VSC CR, Czech Republic; DNRF and DNSRC, Denmark; IN2P3-CNRS, CEA-DRF/IRFU, France; SRNSFG, Georgia; BMBF, HGF, and MPG, Germany; GSRT, Greece; RGC, Hong Kong SAR, China; ISF, I-CORE and Benoziyo Center, Israel; INFN, Italy; MEXT and JSPS, Japan; CNRST, Morocco; NWO, Netherlands; RCN, Norway; MNiSW and NCN, Poland; FCT, Portugal; MNE/IFA, Romania; MES of Russia and NRC KI, Russian Federation; JINR; MESTD, Serbia; MSSR, Slovakia; ARRS and MIZ\v{S}, Slovenia; DST/NRF, South Africa; MINECO, Spain; SRC and Wallenberg Foundation, Sweden; SERI, SNSF and Cantons of Bern and Geneva, Switzerland; MOST, Taiwan; TAEK, Turkey; STFC, United Kingdom; DOE and NSF, United States of America. In addition, individual groups and members have received support from BCKDF, the Canada Council, CANARIE, CRC, Compute Canada, FQRNT, and the Ontario Innovation Trust, Canada; EPLANET, ERC, ERDF, FP7, Horizon 2020 and Marie Sk{\l}odowska-Curie Actions, European Union; Investissements d'Avenir Labex and Idex, ANR, R{\'e}gion Auvergne and Fondation Partager le Savoir, France; DFG and AvH Foundation, Germany; Herakleitos, Thales and Aristeia programmes co-financed by EU-ESF and the Greek NSRF; BSF, GIF and Minerva, Israel; BRF, Norway; CERCA Programme Generalitat de Catalunya, Generalitat Valenciana, Spain; the Royal Society and Leverhulme Trust, United Kingdom.

The crucial computing support from all WLCG partners is acknowledged gratefully, in particular from CERN, the ATLAS Tier-1 facilities at TRIUMF (Canada), NDGF (Denmark, Norway, Sweden), CC-IN2P3 (France), KIT/GridKA (Germany), INFN-CNAF (Italy), NL-T1 (Netherlands), PIC (Spain), ASGC (Taiwan), RAL (UK) and BNL (USA), the Tier-2 facilities worldwide and large non-WLCG resource providers. Major contributors of computing resources are listed in Ref.~\cite{ATL-GEN-PUB-2016-002}.
%----------------------------------------------

%-------------------------------------------------------------------------------

%-------------------------------------------------------------------------------
% If you use biblatex and either biber or bibtex to process the bibliography
% just say \printbibliography here
\clearpage
\printbibliography
% If you want to use the traditional BibTeX you need to use the syntax below.
%\bibliographystyle{bibtex/bst/atlasBibStyleWoTitle}
%\bibliography{compressed_paper,bibtex/bib/ATLAS}
%-------------------------------------------------------------------------------

%-------------------------------------------------------------------------------
\clearpage 
% ATLAS Collaboration author list
% Data extracted on 11-Nov-2017 for paper reference SUSY-2016-25
%\documentclass[11pt]{article}
%\usepackage{a4wide}\begin{document}
\begin{flushleft}
{\Large The ATLAS Collaboration}

\bigskip

M.~Aaboud$^\textrm{\scriptsize 137d}$,
G.~Aad$^\textrm{\scriptsize 88}$,
B.~Abbott$^\textrm{\scriptsize 115}$,
O.~Abdinov$^\textrm{\scriptsize 12}$$^{,*}$,
B.~Abeloos$^\textrm{\scriptsize 119}$,
S.H.~Abidi$^\textrm{\scriptsize 161}$,
O.S.~AbouZeid$^\textrm{\scriptsize 139}$,
N.L.~Abraham$^\textrm{\scriptsize 151}$,
H.~Abramowicz$^\textrm{\scriptsize 155}$,
H.~Abreu$^\textrm{\scriptsize 154}$
% ...
\bigskip
\\
% ...
$^{12}$ Institute of Physics, Azerbaijan Academy of Sciences, Baku, Azerbaijan\\
% ...
$^{88}$ CPPM, Aix-Marseille Universit{\'e} and CNRS/IN2P3, Marseille, France\\
% ...
$^{115}$ Homer L. Dodge Department of Physics and Astronomy, University of Oklahoma, Norman OK, United States of America\\
% ...
$^{119}$ LAL, Univ. Paris-Sud, CNRS/IN2P3, Universit{\'e} Paris-Saclay, Orsay, France\\
% ...
$^{137}$ $^{(a)}$ Facult{\'e} des Sciences Ain Chock, R{\'e}seau Universitaire de Physique des Hautes Energies - Universit{\'e} Hassan II, Casablanca; $^{(b)}$ Centre National de l'Energie des Sciences Techniques Nucleaires, Rabat; $^{(c)}$ Facult{\'e} des Sciences Semlalia, Universit{\'e} Cadi Ayyad, LPHEA-Marrakech; $^{(d)}$ Facult{\'e} des Sciences, Universit{\'e} Mohamed Premier and LPTPM, Oujda; $^{(e)}$ Facult{\'e} des sciences, Universit{\'e} Mohammed V, Rabat, Morocco\\
% ...
$^{139}$ Santa Cruz Institute for Particle Physics, University of California Santa Cruz, Santa Cruz CA, United States of America\\
% ...
$^{151}$ Department of Physics and Astronomy, University of Sussex, Brighton, United Kingdom\\
% ...
$^{154}$ Department of Physics, Technion: Israel Institute of Technology, Haifa, Israel\\
$^{155}$ Raymond and Beverly Sackler School of Physics and Astronomy, Tel Aviv University, Tel Aviv, Israel\\
% ...
$^{161}$ Department of Physics, University of Toronto, Toronto ON, Canada\\
% ...
$^{a}$ Also at Department of Physics, King's College London, London, United Kingdom\\
$^{b}$ Also at Institute of Physics, Azerbaijan Academy of Sciences, Baku, Azerbaijan\\
$^{c}$ Also at Novosibirsk State University, Novosibirsk, Russia\\
$^{d}$ Also at TRIUMF, Vancouver BC, Canada\\
$^{e}$ Also at Department of Physics {\&} Astronomy, University of Louisville, Louisville, KY, United States of America\\
% ...
$^{*}$ Deceased
\end{flushleft}

%\end{document}
% Created with xml2latex.py

%-------------------------------------------------------------------------------

%-------------------------------------------------------------------------------

\end{document}